# Tuning the electron injection mechanism by changing the adsorption mode: the case study of Alizarin on TiO$_2$

Federico Soria, Chiara Daldossi, Cristiana Di Valentin[*]

Dipartimento di Scienza dei Materiali, Università di Milano Bicocca
Via R. Cozzi 55, 20125 Milano Italy

**Abstract**

Functionalized TiO$_2$ nanoparticles with intense fluorescent dyes is a promising tool for several technological applications ranging from photochemistry, photocatalysis, photovoltaics, photodynamic therapy or bioimaging. Here, we present the case study of the Alizarin adsorption on TiO$_2$ nanoparticles (NPs) of different shape and increasing size up to 2.2 nm (700 atoms), by means of density functional theory (DFT) calculations. We find that Alizarin can bind in three different ways, depending on the number and type of bonds between Alizarin and TiO$_2$: "tridented", "bidented" and "chelated". Next, we investigate the optical properties of these systems by time-dependent density functional theory (TDDFT) calculations. Based on the absorption spectra and the Kohn-Sham orbitals analysis, we discovered that the mechanism of electron injection depends on the Alizarin binding mode to the TiO$_2$ surface. While for bidented and chelated adsorption modes a *direct charge transfer* is observed, for the tridented one an *indirect mechanism* governs the charge transfer process following the excitation. Our results are in good agreement with existing experimental data and suggests that by tailoring the shape of the TiO$_2$ NPs and, thus, determining the type of undercoordinated Ti atoms prevalently exposed at the surface, it is possible to control the predominant injection mechanism.

---

[*] Corresponding author: cristiana.divalentin@unimib.it

# 1. Introduction

Titanium dioxide (TiO$_2$) is characterized by outstanding photoabsorption properties. Therefore, TiO$_2$ has been extensively used in photoenergy conversion applications, such as photocatalysis [1–4] photovoltaics [5–9], photoelectrochemistry [10,11] and photochromic devices [12]. Recently, the possibility to exploit the photocatalytic properties of TiO$_2$ in nanomedicine has emerged [13], since the photoexcited hole and electron in TiO$_2$ nanoparticles (NPs) can generate reactive oxygen species (ROS), under UV light [14,15], which can induce cell death [16,17]. As the TiO$_2$ NPs are cytotoxic only once they are light-activated, it is crucial to selectively kill only the targeted sick cells, limiting the side effects on the rest of the body and making them good candidates for cancer treatment [18].

Among the different TiO$_2$ phases, anatase is usually considered to be more suitable for photocatalytic and photovoltaic applications with respect to rutile because anatase is characterized by an indirect band gap of 3.2 eV, which is higher than for rutile (3.03 eV), leading to a slower recombination rate of the photogenerated charge carriers. The major drawback in the use of anatase in photo-applications is that, due to its wide band gap, it only absorbs in the ultraviolet energy region limiting the use of bare TiO$_2$ in photocatalytic and photovoltaic applications. Even more so, the use of TiO$_2$ is restricted in photodynamic therapies since the UV-light required to excite TiO$_2$ does not match the optical window of the biological tissues (600-1000 nm) [13].

Surface coating of TiO$_2$ nanoparticles is used to avoid NPs agglomeration and improve biocompatibility. Surface functionalization can also be used to narrow the semiconductor band gap modifying its range of absorption [19,20]. The surface binding of organic molecules has a significant impact on charge separation, transport, and recombination processes. Once functionalized, TiO$_2$ nanostructures develop the ability to harvest a major portion of the visible and near-infrared spectrum, becoming suitable for photon energy conversion, both for nanomedical [20] and environmental applications [21].

Alizarin is an organic molecule (**Scheme 1**) that has been investigated as a photosensitizer and as a fluorophore [22–29]. The experimental electronic absorption spectrum for free Alizarin shows a low energy band centered at 2.88 eV (431 nm), another shoulder at 3.82 eV (325 nm), and more intense bands at energies higher than 5 eV (250 nm). When Alizarin is adsorbed on TiO$_2$, the lowest band is red shifted by about 0.4 eV and appears at 2.47 eV (503 nm). The other bands are also shifted and appear at energies higher than 3.55 eV (350 nm) [24].

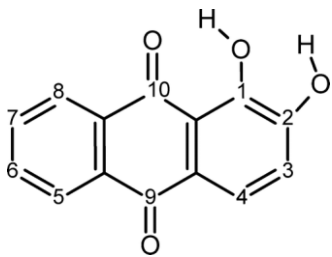

**Scheme 1:** Molecular structure of Alizarin with numbering of the C atoms.

The attachment of Alizarin to TiO$_2$ surfaces has been investigated through resonance Raman spectroscopy experiments [30]. Based on the differences between the free Alizarin spectrum and the Alizarin/TiO$_2$ complex spectrum the authors have proposed a chemisorption of the Alizarin to the TiO$_2$ nanoparticle, probably via the hydroxyl group and/or the carbonyl [30].

When the surface-functionalized TiO$_2$ nanosystems are properly irradiated, an electron is excited from the molecular moiety to the semiconducting oxide. The electron transfer from the HOMO of an adsorbed molecule to the conduction band (CB) of a semiconductor has been modeled in the literature through two different limit mechanisms [31,32], depending on the electron injection mechanism from the molecule to the semiconductor. The first mechanism (type I) involves photoexcitation to a molecule's excited state, from which an electron is then transferred to the solid. The second one (type II) is a direct mechanism, a "one-step" electron injection from the ground state of the adsorbed molecule to the conduction band of the semiconductor. The direct injection mechanism is related to a new charge-transfer band that appears in the absorption spectrum of the molecule upon adsorption, whereas no new bands are seen in the spectrum in the case of the indirect mechanism.

From a theoretical point of view, time-dependent density functional theory (TDDFT) [33] has been used to describe the optical properties of pristine [34] and modified TiO$_2$ systems [31,32,35–37]. Through the comparison of the spectra for free and adsorbed dyes on TiO$_2$, together with the analysis of the molecular orbitals, the authors proposed either a direct or an indirect mechanism for the electron injection and even suggested the existence of possible intermediate mechanisms [36]. Moreover, it was shown that the same dye can present type I or II charge injection mechanisms upon varying the dye conjugation in the anchoring group [31].

In previous theoretical works the spectra of free and adsorbed (on $TiO_2$) Alizarin were calculated using TDDFT [24,32,36,38,39]. Some controversies arose from these works, such as, for example, while the works by Sánchez-de-Armas et. al. [32,36,38] suggest an indirect injection mechanism when Alizarin binds to $TiO_2$ clusters, for Guo et. al. [39] a direct optical charge transfer from Alizarin to the $TiO_2$ cluster is evidenced. Also, from an experimental point of view the results are quite different: an adiabatic electron transfer in Alizarin/$TiO_2$ complex with a 6 fs of time constant was reported by Huber et. al. [28]. However, in another work, Kaniyankandy et. al. [29] concluded that finite size effects of the small NP (~3.4 nm) produce a multiple electron injection with different time scales, which is characteristic of a nonadiabatic event.

In this work, we present a density functional theory (DFT) and a TDDFT investigation of Alizarin-modified $TiO_2$ clusters and a realistic NP of 700 atoms (diameter size of 2.2 nm). First, we have studied the different adsorption modes of the Alizarin molecule on the clusters. We have found that Alizarin can bind to the clusters in three different ways: chelated, bidented, and tridented. In the last binding mode, the chemisorption is through the two O atoms of the hydroxyl groups and the oxygen of a carbonyl group. Then, a study of the optical properties of each Alizarin/$TiO_2$ complex was carried out. We found that depending on the adsorption mode the mechanism of injection changes: for chelated and bidented adsorption a direct injection mechanism is observed whereas an indirect mechanism governs the injection in the tridented adsorption geometry.

## 2. Computational details

The calculations performed in this work were carried out on different models of $TiO_2$ and are based on two levels of theory: density functional theory (DFT) and time-dependent density functional theory (TDDFT). The first has been employed for geometry optimization and electronic structure calculations, whereas the second for simulation of the optical spectra.

### 2.1 *Models.*

In this study, we use three different models of anatase $TiO_2$ (see **Figure 1**): a small cluster of six units of $TiO_2$ (($TiO_2)_6$), namely $Cl_6$, an intermediate cluster of forty units of $TiO_2$ (($TiO_2)_{40} \cdot 2H_2O$), namely $Cl_{40}$, which was built by cutting a finite biatomic layer from an anatase (101) surface supercell and saturating with two water molecules. Finally, we consider a realistic spherical NP with a 2.2 nm diameter, which was previously built by our group through global optimization via a simulated annealing process at the self-consistent charge density functional tight binding SCC-DFTB level of theory and full hybrid DFT (B3LYP) optimization [40]. The stoichiometry of the NP is $(TiO_2)_{223} \cdot 10H_2O$. Even though the $(TiO_2)_6$ cluster was previously shown to be the smallest nanocluster model able to simulate semiquantitatively all the features in the electronic structure and optical properties of real systems [38], the possibility to bind molecules in different configurations (as it would be on real NPs) is reduced. For this reason, it is necessary to use larger or more realistic models.

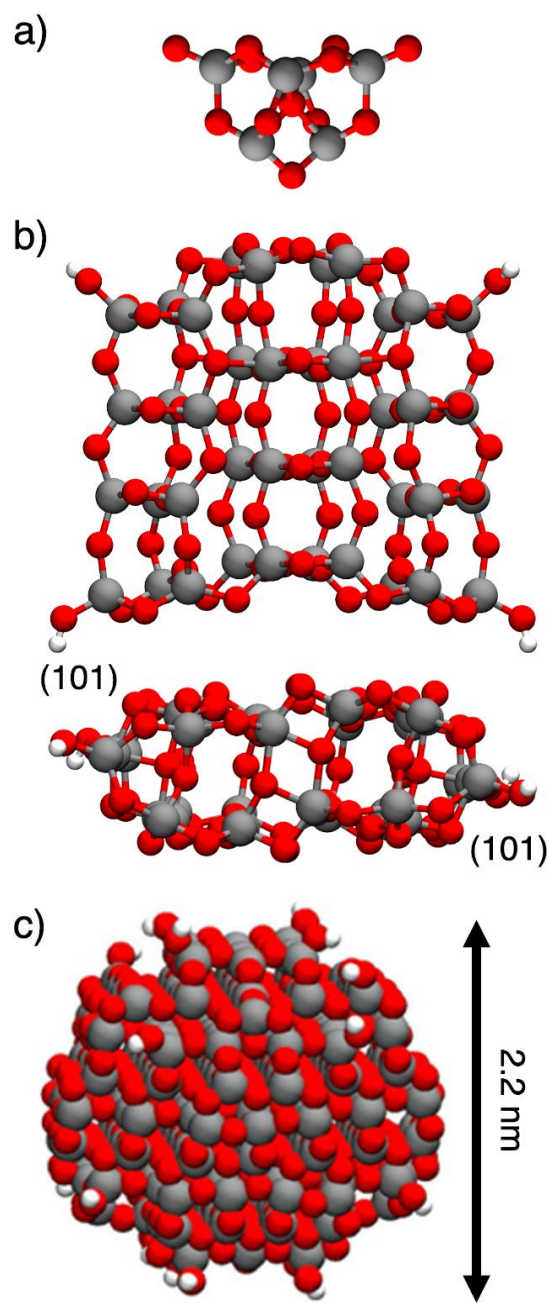

**Figure 1**: Models of the clusters and a realistic NP of TiO$_2$ used in the present work. a) Cluster of six units of TiO$_2$ (Cl$_6$). b) Cluster of forty units of TiO$_2$ (Cl$_{40}$) exposing mainly two anatase 101 facets. c) Spherical realistic NP of anatase with diameter of 2.2 nm.

## 2.2 DFT calculations.

For the DFT optimization calculations, we used the Gaussian16 suite [41] for the small systems ($Cl_6$ cluster) and CRYSTAL14 package [42] for the bigger systems ($Cl_{40}$ cluster and realistic 2.2 nm NP), where the Kohn–Sham orbitals are expanded in Gaussian-type orbitals. For calculations carried out on the $Cl_6$ and $Cl_{40}$ clusters, both PBE [43] and B3LYP [44] functionals were used, while for the 2.2 nm NP only PBE [43]. For Gaussian calculations, a 6–311++G** basis set was used, whereas for Crystal calculations the all-electron basis sets used are: Ti 86–411 (d41), O of $TiO_2$ 8–411 (d1); H 5–11(p1), C 6–311 (d11), O 8–411 (d1) for hydrogen, carbon and oxygen atoms of the Alizarin molecule. The cut-off limits in the evaluation of Coulomb and exchange series/sums appearing in the SCF equation were set to $10^{-7}$ for Coulomb overlap tolerance, $10^{-7}$ for Coulomb penetration tolerance, $10^{-7}$ for exchange overlap tolerance, $10^{-7}$ for exchange pseudo-overlap in the direct space, and $10^{-14}$ for exchange pseudo-overlap. The condition for the SCF convergence was set to $10^{-6}$ au on the total energy difference between two subsequent cycles. The equilibrium structure of isolated moieties and complexes were determined by using a quasi-Newton algorithm with a BFGS Hessian updating scheme [45]. Geometry optimization was performed without any symmetry constraint; forces were relaxed to be less than $4.5 \times 10^{-4}$ au, and displacements to be less than $1.8 \times 10^{-3}$ au.

The adsorption energy of Alizarin on each cluster has been defined as:

$$E_{ads} = [E_{cluster+Alizarin} - (E_{cluster} + E_{Alizarin})]$$

## 2.3 TDDFT calculations.

The electronic absorption spectra were simulated from TDDFT calculations through two sets of complementary calculations. In both cases we have performed the conventional linear response TDDFT (LR-TDDFT) calculations. The Gaussian16 suite [41] was used for the isolated and functionalized $(TiO_2)_6$ and $(TiO_2)_{40} \cdot 2H_2O$ clusters using PBE [43] and B3LYP [44] functionals. For the small systems we have also performed calculations using the hybrid PBE0 [46] and the range-separated CAM-B3LYP [47] functionals. For the 2.2 nm NP, due to the size and the number of atoms, the LR-TDDFT calculation was carried out using only the PBE functional and the Octopus code [48], in which the electron-nucleus interaction is described within the pseudopotential approximation. For the $Cl_6$ system we computed 300 transitions, for the $Cl_{40}$ and NP systems 100 transitions.

## 3. Results and Discussion

### 3.1 *Adsorption modes of Alizarin on TiO$_2$*

As detailed above, the adsorption of Alizarin on the TiO$_2$ was studied on three different models. It is worth mentioning that previous theoretical TDDFT studies [24,32,36,38,39] on Alizarin binding TiO$_2$ clusters only considered a chelated mode to investigate the optical properties. This choice was based on the observation that a similar molecule, namely catechol, according to Solid-State NMR experiments combined with DFT calculations, adsorbs on TiO$_2$ mainly in a chelated binding mode, with some degree of bidented attachment [49]. For these reasons, subsequently Alizarin was proposed to bind in a similar way [38]. However, the chemical structure of Alizarin is different to the catechol due to the presence of the carbonyl groups, which can interact with the uncoordinated Ti atoms on a surface. Raman spectroscopy experiments suggest that Alizarin attaches to the TiO$_2$ surface via the hydroxyl and/or the carbonyl groups [30]. **Figure 2** shows the optimized structures of the Alizarin adsorbed on the Cl$_6$ and Cl$_{40}$ clusters and on the 2.2 nm NP. In all these cases we study the dissociative adsorption of Alizarin. It means that when the molecule is adsorbed on TiO$_2$ the H atoms initially in OH groups of the Alizarin migrate to two different surface O atoms of the TiO$_2$. In this way the system remains neutral. Charged systems will be studied in Section 3.4.3.

On the Cl$_6$, we found that the Alizarin can link in two different modes (**Figure 2 and Figure S1**): the "so-called" chelated, where the two O atoms of the hydroxyl group (which dissociate) bond to the same uncoordinated Ti$_{4C}$ or Ti$_{3C}$ atoms [39]. The adsorption energy with the PBE functional for this conformation is -2.38 eV. On the other hand, we found that the chelated mode becomes a 'tridented' one when a carbonyl O atom of the Alizarin binds to another Ti atom of the cluster. In this case the adsorption energy is E$_{ads}$ = -2.99 eV indicating that tridented configuration is the most stable. The same trend was observed with both PBE or hybrid B3LYP functionals (**Table 1**).

On the Cl$_{40}$ cluster and on the NP we found three binding modes: chelated and tridented, as in Cl$_6$, but also a bidented one. This is possible because both in the Cl$_{40}$ and in the NP there are two neighboring Ti$_{5C}$ atoms, which are not present in the small cluster. **Figure 2** and **Figure S2** show the configurations and details of each structure, respectively. On the Cl$_{40}$ the adsorption energies with the PBE functional are -0.18, -0.43 and +0.14 eV for the tridented, bidented and chelated modes, respectively. In line with what observed for Cl$_6$, also in the case of $Cl_{40}$ the B3LYP functional provides the same trend of adsorption as the PBE functional: the bidented is the most stable configuration followed by the tridented and chelated ones.

Finally, on the realistic NP we found that Alizarin can adsorb in two tridented modes, in a bidented and in a chelated one. The difference between the two tridented modes lies in the choice of O atoms of the Alizarin binding the molecule to the surface. While in one structure the two O atoms of the hydroxyl groups are bonded to the same $Ti_{4c}$ atom and the O atom of the carbonyl group binds to another surface $Ti_{4c}$ atom (**Figure 2**), in the other structure the O atom of one hydroxyl group and that of the carbonyl group are bonded to the same Ti atom and the O atom of the other hydroxyl group binds to a different Ti atom (see **Figure S3** for details). For the two configurations the adsorption energies with PBE functional are -3.48 and -3.32 eV, respectively. As in the $Cl_{40}$ the bidented mode involves two $T_{5c}$ atoms on the NP, whereas the chelated mode occurs when the Alizarin binds to a single $Ti_{4c}$ atom. The adsorption energies for these configurations are -1.46 and -2.12 eV, respectively. Comparing the adsorption energies computed in the case of the NP (**Table 1**), the tridented mode is the most stable, as previously observed for the small $Cl_6$ cluster. It is worth noting that the bidented adsorption mode cannot involve two $Ti_{4c}$ atoms. During the atomic relaxation starting from a bidented geometry, the Alizarin/$TiO_2$ evolves into the tridented optimized structure. This behavior is different to what observed for Dopamine or DOPAC molecules adsorbed on the same NP, where the bidented way of adsorption is observed on two $Ti_{4c}$ surface atoms [50]. The difference is because the Dopamine and DOPAC do not have an extra O atom, which can further bind to one of the $Ti_{4c}$ atoms.

Based on data in **Table 1**, we notice that the adsorption energies for $Cl_{40}$ are much smaller than those for $Cl_6$ or for the NP and that the chelated mode is even energetically unstable. One possible reason could be that $Cl_{40}$ resembles a portion of a flat anatase (101) surface, which is known to be less reactive than nanoparticles [51]. However, we thought that this discrepancy required further attention and decided that, to achieve a deeper understanding of this different behavior of $Cl_{40}$, some extra calculations were necessary. First, we calculated the adsorption energies considering the dispersion contribution in the calculations using the Grimme approach [52,53]. When we include this correction, we found that the values of $E_{ads}$ on $Cl_{40}$ become more negative, i.e. -0.73, -1.12 and -0.16 eV for tridented, bidented and chelated modes, respectively. Analogously, for $Cl_6$ the adsorbed species are further stabilized with $E_{ads}$ of -2.70 and -3.70 eV for chelated and tridented modes, respectively. Therefore, since the energy gain is similar, the difference in the magnitude of the adsorption energies between $Cl_{40}$ and $Cl_6$ is maintained, even after including the dispersion contributions. In order to understand this net difference, we carried out a decomposition of the adsorption energies. **Table S1** in the Supp. Mater. reports the adsorption energy, the cluster and Alizarin deformation energies and the binding energy for the various adsorption modes on the different $TiO_2$ cluster/NP models, using the PBE functional. We observe that the deformation energies of the $TiO_2$ model upon Alizarin adsorption is higher for $Cl_{40}$ than for the 2.2 nm NP. For instance, for the tridented mode, the deformation energy of the $Cl_{40}$ cluster is 6.12 eV, while the deformation energy of the 2.2 nm NP is 3.91 eV. The same trend holds for the bidented and chelated adsorption modes, too (see **Table S1**). The binding energy is evaluated as the energy difference between the complex and the separated fragments (Alizarin and cluster/NP) but keeping them in the same configuration as in the complex (see **Table S1**). We notice that the binding energies for $Cl_{40}$ and the 2.2 nm NP are similar when comparing the same binding mode. This implies that: a) the bonding extent for $Cl_{40}$ or for the 2.2 nm NP is similar and b) the difference in the adsorption energy ($E_{ads}$) is mostly due to a higher distortion of $Cl_{40}$ with respect to the NP upon Alizarin adsorption. We found a linear correlation between the difference in the adsorption energy and the difference in the deformation energy for each configuration, between $Cl_{40}$ and NP (see **Figure S4**). Regarding the Alizarin/$Cl_6$ complexes, we observe large adsorption energy values, like those computed for the NP, although the deformation energies are similar or larger than for the NP. The reason of this peculiar behavior is the following: the O atoms of $Cl_6$ accepted the protons coming from the Alizarin are monocoordinated (and more reactive, i.e. Ti=O) in contrast to those in the NP that are all twofold coordinated.

In summary, we found that Alizarin can bind to the $TiO_2$ nanostructures in many different modes either by the hydroxyl or the carbonyl groups. We proved that the tridented mode, which has never been investigated before, is the most stable configuration on the $Cl_6$ and on the 2.2 nm NP. On the $Cl_{40}$ cluster, however, the bidented mode is the most stable configuration due to a lower $Cl_{40}$ deformation when Alizarin is bound. As we will show in the next sections, the different adsorption modes provide different injection charge mechanisms, therefore it is crucial to determine which type of surface one is dealing with.

**Table 1**. Calculated adsorption energies in eV for the adsorption of Alizarin in the different configurations on the $Cl_6$ and $Cl_{40}$ $TiO_2$ models with PBE and (B3LYP) functional and on a realistic NP with PBE functional.

|  | Tridented | Bidented | Chelated |
|---|---|---|---|
| $Cl_6$ | -2.99(-3.15) | - | -2.38(-2.53) |
| $Cl_{40}$ | -0.18(-0.05) | -0.43(-0.58) | +0.14(+0.04) |
| NP | -3.48/-3.32 | -1.46 | -2.12 |

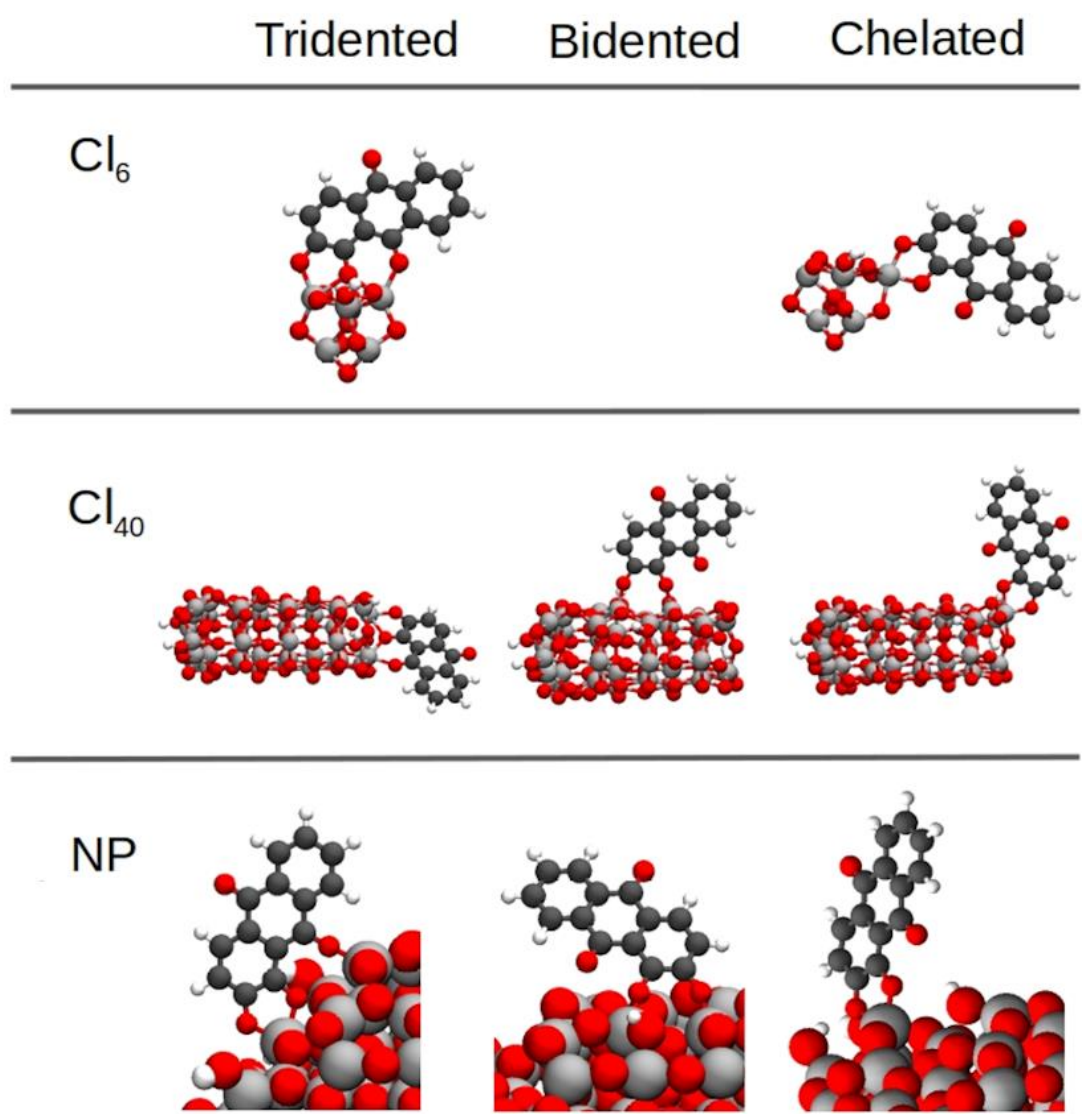

**Figure 2**: Optimized structures of the adsorption of Alizarin in tridented, bidented and chelated modes on a) $Cl_6$ cluster, b) $Cl_{40}$ cluster and c) the 2.2 nm NP.

## 3.2 *Alizarin absorption spectrum*

As we are interested in the change of the absorption spectra when Alizarin binds onto a $TiO_2$ surface, first we obtained the absorption spectrum of isolated Alizarin in different media. The results are reported in **Figure 3** and **Table 2.** Using PBE functional the lowest absorption peak appears at 2.31 eV (536 nm) (black line in **Figure 3a**), while using the B3LYP functional it is at 2.86 eV (433 nm) (blue line in **Figure 3a**). In both calculations, the first absorption peak corresponds to a single HOMO → LUMO transition (**Figure 3b**). Compared with experimental data [54,55], B3LYP better reproduces the position of the first absorption band. Our calculation is also in line with previous ones [56,57]. We tested other DFT functionals, such as PBE0 and the range-separated CAM-B3LYP (see **Figure S5,** and **Table S2**). While PBE0 results are similar to B3LYP ones, CAM-B3LYP overestimates the position of the first band by 0.6 eV. We also considered the inclusion of the solvent effects by means of the PCM model [58]. For these calculations, we used the B3LYP functional and relaxed the Alizarin geometry in the presence of the implicit solvent (water and benzene). An enhancement of the absorption strength with respect to the molecule in vacuum is observed in agreement with previous works [56]. The first peak does not shift in energy when PCM is applied (see **Figure 3c**), while the second and third absorption peaks show a slight red-shift around 0.1 eV with respect to the vacuum conditions. It is interesting to note that in both water and benzene, although an enhancement of absorption and some red-shift is observed, there is no change in the overall shape of the spectrum.

**Table 2:** Lowest absorption peak energy (eV) and HOMO-LUMO gap (eV) for free Alizarin using the PBE and B3LYP functionals, obtained from LR-TDDFT and DFT calculations respectively, with and without inclusion of the solvent effect (water and benzene) through the PCM model. Experimental value for the first absorption band is also reported in eV.

|  | PBE /vacuum | B3LYP /vacuum | B3LYP /water | B3LYP /benzene | Exp. |
|---|---|---|---|---|---|
| First excitation | 2.31 | 2.86 | 2.83 | 2.82 | 2.88 |
| HOMO-LUMO gap | 1.88 | 3.30 | 3.29 | 3.30 |  |

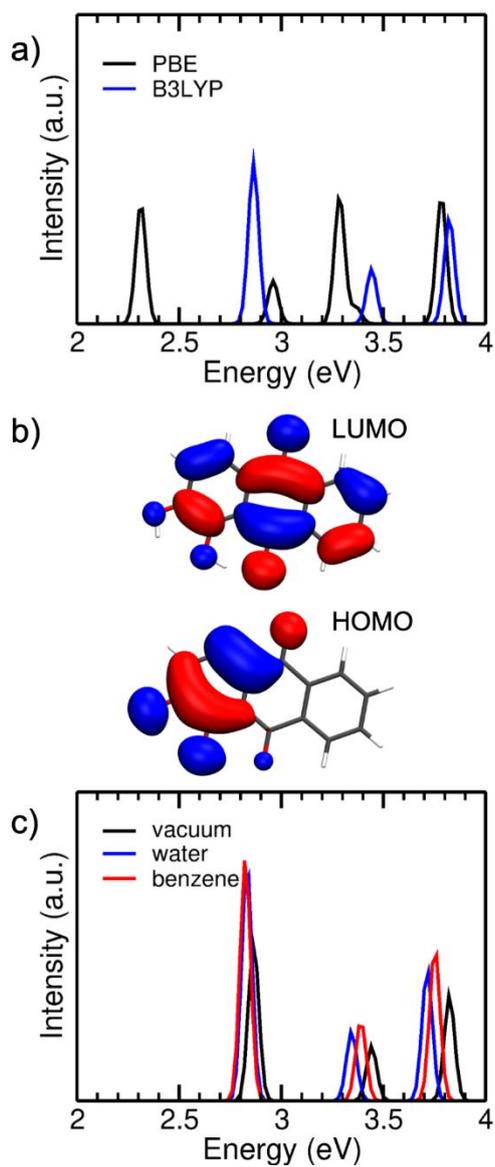

**Figure 3**: a) Alizarin spectra calculated in vacuum using B3LYP and PBE functionals. b) HOMO and LUMO molecular orbitals of Alizarin. c) Alizarin spectra calculated in vacuum, in water and in benzene solvents using B3LYP functional.

### 3.3 *Absorption spectra of TiO$_2$ model nanostructures*

In this section we investigate the absorption spectra of the different TiO$_2$ clusters and of the 2.2 nm NP. **Figure 4a** shows the spectra of the Cl$_6$ and Cl$_{40}$ clusters and of the 2.2 nm NP calculated using the PBE functional, whereas in **Figure 4b** the spectra of the Cl$_6$ and Cl$_{40}$ using B3LYP functional is reported. It is possible to note that using the PBE functional, the first band is centered in 2.68 and 2.63 eV for Cl$_6$ and Cl$_{40}$, respectively (black and blue lines in **Figure 4a**). For both clusters, the TDDFT absorption threshold corresponds well with the computed HOMO-LUMO gaps computed with standard GGA functionals (2.66 and 2.63 eV respectively) (see **Table 3**). Using the hybrid functional B3LYP, the first excitations are at 4.01 and 3.85 eV for Cl$_6$ and Cl$_{40}$, respectively. In the case of the B3LYP functional, TDDFT absorption onsets are 0.7 and 0.6 eV lower than the corresponding HOMO-LUMO gaps, which is similar to what previously reported [59]. For the 2.2 nm NP, due to the size and the number of atoms, the calculation was carried out with the less computationally costly PBE functional and with the Octopus code (for details see the Section 2.3 TDDFT calculations). The spectrum shows a first excitation band at 2.45 eV (green line in **Figure 4a**), which well corresponds to the HOMO-LUMO gap of 2.45 eV (see **Table 3**). It is worth noting how the optical gap lowers from 2.68 to 2.45 eV going from Cl$_6$ to the realistic 2.2 nm NP, which shows a progressive convergence of the gap towards the bulk value of 2.12 eV [60], using the same PBE functional.

As we did for the Alizarin molecule, we now compare the PBE results for the Cl$_6$ with those obtained with the hybrid PBE0 functional, the range-separated CAM-B3LYP functional and the inclusion of solvent through the PCM model. **Figure S6** and **Table S3** show that the PBE0 and CAM-B3LYP methods provide absorption onsets for the Cl$_6$ of 4.39 and 4.75 eV, respectively, in agreement with previous reports [61]. Regarding the solvent effect, we observe a blue-shift of the first excitation energies from 4.01 in vacuum to 4.16 and 4.38 eV in benzene and water, respectively (**Figure S7** and **Table S4**).

**Table 3** Lowest band energies and HOMO-LUMO gap (eV) for the Cl$_6$, Cl$_{40}$ and 2.2 nm NP obtained from LR-TDDFT and DFT calculations respectively.

|                 | Cl$_6$ |       | Cl$_{40}$ |       | 2.2 nm NP |
| --------------- | ------ | ----- | --------- | ----- | --------- |
|                 | PBE    | B3LYP | PBE       | B3LYP | PBE       |
| First excitation | 2.68  | 4.01  | 2.63      | 3.85  | 2.45      |
| HOMO-LUMO gap   | 2.66   | 4.70  | 2.63      | 4.44  | 2.45      |

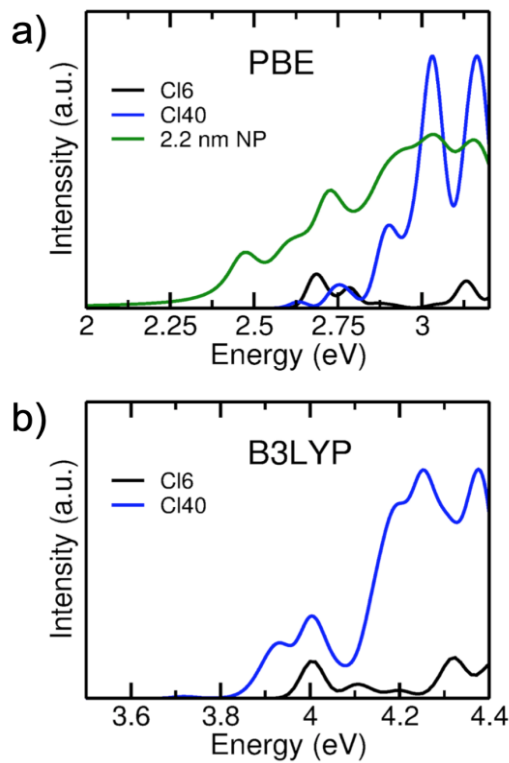

**Figure 4**: a) Spectra of $Cl_6$, $Cl_{40}$ and 2.2 nm NP calculated with the PBE functional. b) spectra of $Cl_6$ and $Cl_{40}$ calculated by the use of B3LYP functional.

Based on the behavior of the different functionals in the description of both Alizarin and the $TiO_2$ clusters, we can conclude that the one which best resembles the experimental observations is the B3LYP hybrid functional. Thus, it is the method of choice to accurately describe the charge injection mechanism.

### 3.4 *Absorption spectra of Alizarin/TiO$_2$ complexes*

In this section we investigate how the absorption spectrum of Alizarin changes when the molecule binds to a $TiO_2$ nanostructure. We consider all the complexes shown in **Figure 2**. As aforementioned, previous works considered only the chelated adsorption mode [24,32,36,38,39]. However, we have shown above that this configuration is not the most stable (see **Table 1**), but, on the contrary, the bidented or tridented adsorption mode have larger adsorption energies, depending on the $TiO_2$ models considered. For this reason, in the following we will present the optical properties of all the three binding configurations. For $Cl_6$ and $Cl_{40}$ we will discuss the results obtained using the B3LYP functional, which was proven to be the most accurate for both the Alizarin molecule and for the $TiO_2$ models in Sections 3.1 and 3.2. Based on the validation of the PBE method against B3LYP results on $Cl_6$ and $Cl_{40}$, as presented in Supp. Mater., only optical spectra obtained with PBE functional will be presented in the case of the Alizarin/NP complexes, due to the too large computational cost of using B3LYP functional for systems composed of more than 700 atoms.

### 3.4.1 Alizarin/Cl$_6$ complex

In **Figure 5a** we present the absorption spectra, obtained with the B3LYP functional, of the Alizarin/TiO$_2$ complexes for the Cl$_6$ cluster and compare them with free Alizarin. In the tridented and chelated modes, the first adsorption band is found at 2.03 and 2.08 eV respectively. Compared with the spectrum of isolated Alizarin, a clear red-shift in the absorption threshold of 0.83 and 0.78 eV, respectively, is observed. However, we will show that the origin and nature of this red-shift is different depending on the adsorption mode.

In **Table S5** the assignment of the electronic excitations for tridented and chelated adsorption modes are displayed. For the tridented case, the first band is made up mainly by the HOMO → LUMO transition. Based on the analysis of the Kohn Sham orbitals energies (**Figure 5b**) it is possible to note that while the HOMO orbital is almost at the same position as the free Alizarin, the LUMO is stabilized by -0.67 eV with respect to the free Alizarin. The inspection of the molecular orbital plots (**Figures 5c and S8**) suggests that both HOMO and LUMO are similar to those of the free Alizarin (**Figure 3b**). Thus, for the tridented adsorption mode the first band in the optical spectrum corresponds to a red-shift of the HOMO-LUMO transition in the free Alizarin, which is a clear indication of an indirect mechanism of charge injection.

On the other hand, the analysis of the chelated mode shows that the first intense peak at 2.08 eV consists of four transitions, of which the most important ones being HOMO→LUMO+3 and HOMO→LUMO+4 (**Table S5**). These molecular orbitals are mainly localized on the cluster as can be seen in **Figure S8.** Even LUMO, LUMO+1 and LUMO+2 orbitals, which are involved in the small feature at 1.93 eV in the absorption spectrum, are localized on the cluster **(Figures 5d and S8)**. The Kohn Sham orbitals energies show that the HOMO orbital in the chelated configuration is less stable than in the free Alizarin whereas the LUMO orbital of the whole system is lower in energy with respect to the LUMO for free Alizarin. However, this virtual state of the complex does not correspond to the LUMO orbital of the free Alizarin. We found that the virtual molecular orbital having similar characteristics to the LUMO orbital of the free Alizarin is the LUMO+9 state (**Figure S8**). This is the main state involved in the transition whose peak is located at 2.75 eV (see **Table S5**). Then, the analysis above reveals that, for the chelated mode, the mechanism of injection is different compared with the tridented one. The first peak in the absorption spectrum of the chelated mode does not correspond to the red-shift of the first band in the absorption spectrum of the free Alizarin. Not even the absorption peaks at 2.21 and 2.52 eV correspond to transitions to virtual orbitals having similar characteristics to the LUMO of the free Alizarin. On the contrary, the virtual states in the excitations at 1.93, 2.08, 2.21 and 2.52 eV are in all cases localized on the cluster, suggesting that, for this type of adsorption, the charge injection mechanism is direct (or type II). Next, we will show that these different behaviors are independent of the cluster size but are only related to the molecular binding mode.

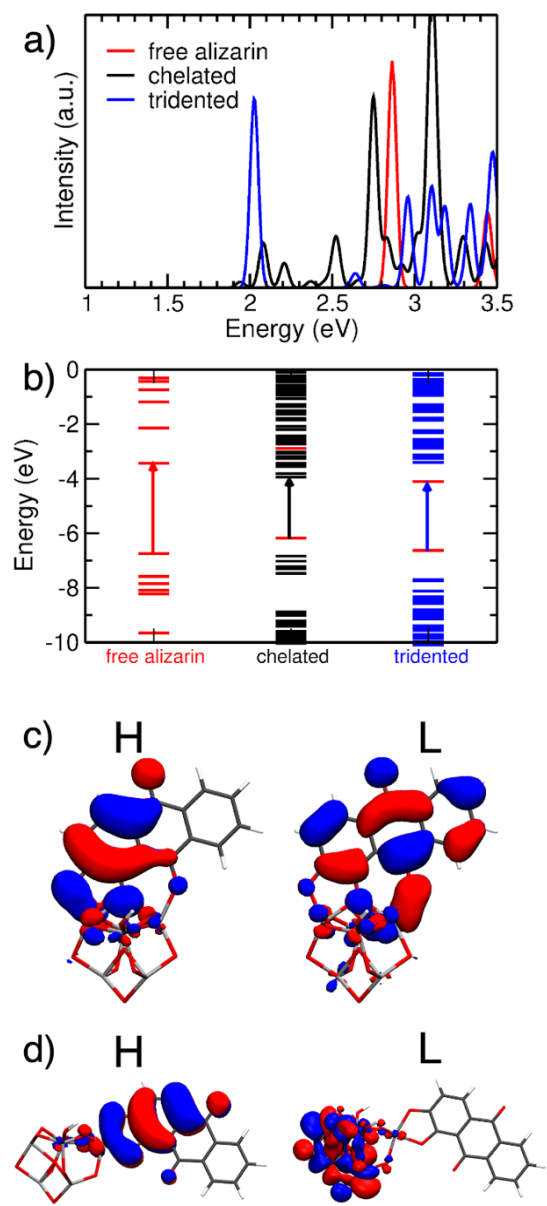

**Figure 5.** a) Absorption spectra (B3LYP) of tridented and chelated complexes on $Cl_6$ together with the Alizarin spectrum. b) Molecular orbital energies for the ground state of free Alizarin, and the complexes. For each complex we have marked the HOMO orbital and the orbital that corresponds to the LUMO of the free molecule. c) and d) HOMO and LUMO orbitals of adsorbed Alizarin in tridented and chelated mode, respectively.

### 3.4.2 Alizarin/$Cl_{40}$ complex

In **Figure 6a** we report the absorption spectra of the Alizarin attached to $Cl_{40}$ in the three different configurations: tridented, chelated and bidented. As for the small $Cl_6$ cluster, many differences can be observed after careful analysis. In the tridented absorption mode the spectrum starts with an intense peak at 1.87 eV, while the spectrum of the chelated one presents some small peaks at low energy between 1.30 and 1.85 eV. Regarding the bidented mode, the absorption spectrum starts with an intense band at 2.39 eV.

The analysis of the molecular orbitals and excitation energies provide similar conclusions to those drawn for the small cluster in the previous section. Data reported in **Table S6** show that the first peak for the tridented configuration is mostly made up by the HOMO→ LUMO transition. As for $Cl_6$ case, it is possible to observe a stabilization of 0.78 eV of the LUMO orbital with respect to that of the free Alizarin (see **Figure 6b**). Plots in **Figures 6c and S9** show that the HOMO orbital is quite like the HOMO orbital of the free Alizarin and the LUMO orbital, even though there is some contribution from the $TiO_2$ cluster, is mostly a molecular orbital with the same characteristics of the LUMO of free Alizarin. These results suggest that for this mode of adsorption the mechanism of injection is indirect.

Differently, the absorption spectrum of the chelated configuration shows some small peaks at the beginning that have contributions from many transitions, as is shown in **Table S6.** In these transitions the excited states are always centered on the cluster. In **Figure 6b** the Kohn Sham orbitals diagram shows that the HOMO orbital in the chelated configuration is less stable than in the free Alizarin, while the LUMO orbital of the whole system is stabilized with respect to that of free Alizarin. However, as for $Cl_6$, this virtual state does not correspond to the LUMO orbital of the free Alizarin (**Figure 6d**). The virtual molecular orbital having similar characteristics to the LUMO orbital of the free Alizarin is the LUMO+86 state (**Figure S9**), which is found at much higher energies. Clearly there are many states between the LUMO and LUMO+86 orbitals that are responsible for the many transitions observed and reported in **Table S6.** These orbitals are mainly centered on the $TiO_2$ cluster, advising a direct mechanism of charge injection.

Finally, in the bidented adsorption mode we found that the initial intense band at 2.39 eV is made up of two excitations involving seven transitions: the most important are the HOMO→LUMO and HOMO→LUMO+1 for the first excitation and HOMO→LUMO+2 and HOMO→LUMO+3 for the second excitation (see **Table S6**). These four virtual orbitals are centered mainly on the $TiO_2$ cluster, as shown in **Figures 6e and S9**. For the peak at 2.82 eV, the main excitation is made up by eight transitions, of which the two most important are the HOMO-1→LUMO+1 and HOMO→LUMO+17. This last virtual orbital is also centered on the TiO$_2$ cluster. Thus, the analysis suggests that, as for the chelated mode, for the bidented one a direct charge injection mechanism is taking place.

The excited states were also examined with the charge transfer distance index, D$_{CT}$, which measures the spatial extent of charge transfer excitations in Å and the charge passed index, q$_{CT}$, which is the integration of the density depletion function over all space in atomic units, au. [62,63] The index D$_{CT}$ is calculated as the distance between the barycenter of density increment and that of density depletion regions. **Table S7** in the Supp. Mater., reports the values of D$_{CT}$ and q$_{CT}$ for the free gas phase Alizarin and when it is attached on Cl$_6$ and Cl$_{40}$. The D$_{CT}$ and q$_{CT}$ values for Alizarin molecule are 3.55 Å and 0.75 respectively. When the molecule is adsorbed in the tridented mode on both Cl$_6$ and Cl$_{40}$ clusters, the values of D$_{CT}$ and q$_{CT}$ are computed to be very similar to those of a free gas phase molecule: 3.25 Å and 0.70 on Cl$_6$ and 3.68 Å and 0.81 on Cl$_{40}$, respectively. This analogy is a hint for an indirect mechanism, where an excitation between the molecular orbitals of Alizarin is envisaged first. When Alizarin is attached in the chelated mode, the values of D$_{CT}$ and q$_{CT}$ are quite different with respect to the tridented ones: 7.01 Å (Cl$_6$) and 9.32 Å (Cl$_{40}$) for D$_{CT}$ and 0.97 in both cases for q$_{CT}$. This is a large distance in the spatial extent of charge transfer excitations, which can be rationalized with the excited state of the transition being largely localized on the cluster, which confirms a direct charge transfer, as proposed above according to the molecular orbitals analysis. Similar analysis can be done for the bidented mode (on Cl$_{40}$) since again here the values of D$_{CT}$ and q$_{CT}$ are larger than for free gas phase Alizarin, corroborating the direct charge transfer mechanism based on the molecular orbitals analysis discussed above.

In the Supp. Mater., we provide evidence that the different observed mechanisms are independent of the functional used. In **Figure S10** the simulated spectra and the diagram of the molecular orbitals obtained with the PBE functional present the same general characteristics of the different mechanisms discussed above, based on the B3LYP results. Also, in the Supp. Mater. (**Figure S11**), we present a parallel study of the electron injection mechanism using the range separated function CAM-B3LYP, which is known to be accurate in the description of intermolecular electron transfer reactions [61]. This type of calculations was performed only for the Alizarin/Cl$_6$ complexes. The main characteristics that we reported based on B3LYP calculations are consistent with the results by CAM-B3LYP functional: a) the tridented system starts to absorb at lower energy than the chelated one and b) the first excitation for the tridented system is basically the Alizarin HOMO→LUMO transition, whereas for the chelated mode the first transition is a direct charge injection from the molecule to the TiO$_2$ cluster.

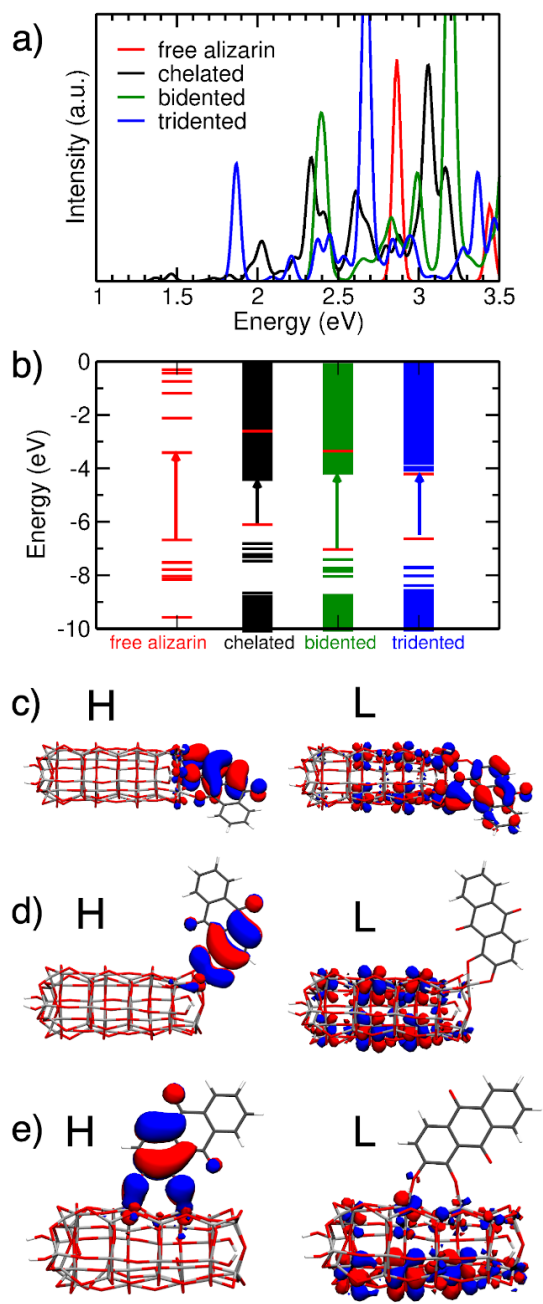

**Figure 6.** a) Absorption spectra (B3LYP) of tridented, chelated and bidented complexes on Cl$_{40}$ together with the Alizarin spectrum. b) Molecular orbital energies for the ground state of free Alizarin, and the complexes. For each complex we have marked the HOMO orbital and the orbital that corresponds to the LUMO of the free molecule. c), d) and e) Selected occupied and virtual molecular orbitals of adsorbed Alizarin in tridented, chelated and bidented mode respectively.

In order to compare our findings directly with previous experimental results, in **Figure 7** we plot again the spectra shown in **Figure 6a** but using the wavelength scale and a larger broadening, together with the experimental spectrum registered for Alizarin-sensitized $TiO_2$ nanoparticles of a diameter of about 16 nm [25]. We observed that the best fit with the experimental spectrum comes from the bidented adsorption model presenting a band around 500 nm. As for the spectrum of the chelated model, a band at ~ 500 nm is also observed, but two shoulders at ~450 nm and at ~640 nm are present, which are not seen in the experimental spectrum. The spectrum of the tridented adsorption model shows a broader band centered at ~660 nm and a sharp and intense band at ~440 nm, which are also not present in the experimental spectrum. The comparison suggests that, in the experiment, Alizarin molecules bind to the $TiO_2$ NP mainly in the bidented mode. We can rationalize this behavior based on availability of the $Ti_{4c}$ and $Ti_{5c}$ sites on the NP surface. We have shown in Section 3.1 that for the adsorption of Alizarin in the tridented mode two $Ti_{4c}$ sites (or one $Ti_{4c}$ and one $Ti_{5c}$) are required, in the chelated mode only one $Ti_{4c}$ is required, while in the bidented mode two $Ti_{5c}$ surface atoms must be available. In a previous work by our group [40], we showed that the % of surface $Ti_{4c}$ sites progressively decreases as the size of the NP increases (diameter going from 1.5 to 4.4 nm) relative to the % of $Ti_{5c}$ sites. On this basis, one can extrapolate that in a NP of 16 nm (as those used in the experiment) the portion of $Ti_{4c}$ surface sites is very low in comparison with that of $Ti_{5c}$. Thus, we conclude that the chelated and tridented adsorption modes are less probable to be found on a large NP of 16 nm. On the contrary, the bidented mode is expected to be the most popular type of adsorption and, therefore, to dictate the shape of the absorption spectrum.

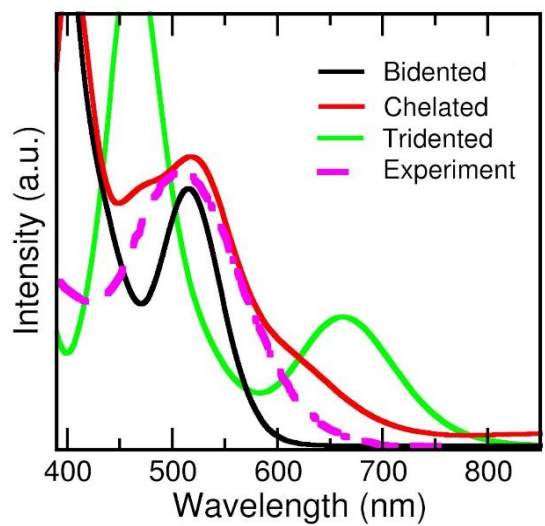

**Figure 7.** Absorption spectra (B3LYP) of bidented, chelated and tridented complexes on Cl$_{40}$ together with the experimental spectrum obtained for Alizarin-sensitized TiO$_2$ nanoparticles with a diameter of about 16 nm (rescaled form Ref. [25]).

### 3.4.3 pH effect on the optical spectra and the injection mechanism

In this section we investigate the pH effect on the absorption spectra on the Alizarin/TiO$_2$ systems and its implication on the electron injection mechanism. Experimentally, the spectrum of the Alizarin/TiO$_2$ changes as the pH goes from acidic to basic conditions [26,64]. The first excitation band at 496 nm is observed at pH lower than 5. In the range of pH 7−12, the absorption band maximum is red-shifted to values between 526 and 535 nm. At pH 5 the Alizarin molecule is expected to be neutral, while increasing the pH produce the formation of monoanionic and dianionic species (pK$_1$ and pK$_2$ values of 6.6 and 12.4 in dioxane/water (1:2) at room temperature) [65]. In order to simulate the adsorption spectra of free and adsorbed Alizarin on TiO$_2$ at different pH conditions, we carried out a set of calculations of singly and doubly negatively charged systems and the results are discussed below.

In the case of a free Alizarin molecule, the calculated lowest peak for the neutral form is 2.86 eV (see **Table 2**), while for the monoanionic/dianionic forms it is 2.00/1.87 eV, respectively. These results are in good agreement with the experimental trend [26,64]. For the anions it is possible to observe a larger destabilization of the HOMO with respect to LUMO (both strongly destabilized, see **Figure S12**), which leads to the observed red-shift in the absorption spectra at increasing pH conditions.

When the neutral or anionic forms of Alizarin are adsorbed on Cl$_6$, we found that the optical spectrum changes as well as the injection mechanism. The first observation is that for the chelated adsorption mode there is a red-shift in the first excitation band of the absorption spectra going from the neutral to the monoanionic and from the monoanionic to the dianionic forms (**Figure 8a**). A similar trend was observed experimentally for Alizarin/$TiO_2$ systems in 60% ethylene glycol in water [26] and aqueous [64] solutions, respectively, at increasing pH conditions. For the tridented system, the position of the first absorption peak does not change going from the neutral to the monoanionic form, while for the dianionic one we observe a small red-shift (**Figure 8b**). In both cases (chelated and tridented adsorption modes), when Alizarin anions (monoanion or dianion) are attached on the surface, a clear stabilization of the molecular orbitals is observed with respect to the free anions. This stabilization can be rationalized as due to a large redistribution of the negative charge over the $TiO_2$ cluster. For larger clusters, this stabilization is expected to be even greater. Next, we comparatively analyze the behavior of adsorbed species with free ones. For the neutral Alizarin this comparison is already present for different adsorption modes in **Section 3.4.1.** Here, we analyze the monoanionic and dianionic species.

For the chelated monoanionic system, we found that the first excitation band at 2.55 eV (485.6 nm) consists of three transitions, of which the most important ones being HOMO→LUMO+2 and HOMO→LUMO+3. The inspection of the molecular orbital plots shows that the HOMO corresponds to the HOMO of the free Alizarin (see **Figure S13** vs **Figure 3**). The LUMO+2 is similar to the LUMO of the free Alizarin, while the LUMO+3 is spread mainly on the $TiO_2$ cluster. This suggests that for the chelated monoanionic system an intermediate mechanism is active.

For the chelated dianionic system, the first band at 2.25 eV (552.2 nm) is mostly made up by the HOMO→LUMO transition. The inspection of the molecular orbital plots shows that both HOMO and LUMO are similar to those of the free Alizarin (**Figure S13** vs **Figure 3**) which is a clear indication of an indirect mechanism of charge injection. This is the opposite of what is observed for the neutral system, for which a direct mechanism was determined.

Thus, for chelated adsorption mode there is a transition from a purely direct mechanism for the neutral system to a purely indirect mechanism for the dianionic system going through an intermediate situation for the monoanionic system.

For the tridented systems, we do not observe a change in the charge injection mechanism from neutral to monoanionic and dianionic systems, which is always indirect. This is confirmed by the fact that in all these cases the first excitation is made of the Alizarin HOMO→LUMO transition (**Figure S14**).

In summary, based on these calculations on anionic species, we can conclude that the electron injection mechanism does not just depend on the mode of adsorption of Alizarin on the $TiO_2$, but also on the pH conditions.

We wish to conclude this analysis by noting that the shift in the first absorption peak going from neutral to monoanionic to dianionic chelated Alizarin (406, 486 and 552 nm, respectively, as shown in **Figure 8**) is quite overestimated with respect to the experimental data in **Figure 2** in Ref. [26] (495, 505 and 512 nm, respectively). We attribute this overestimation to the unscreened charge of the anionic species, which could be solved by inserting the systems in an implicit solvent (water) with a specific dielectric constant, which is more similar to the experimental situation. To prove this hypothesis, we calculated again the absorption spectra of Alizarin/$Cl_6$ in water using the PCM model. Indeed, the shift decrease as follow: 456, 482 and 500 nm, for neutral, monoanionic and dianionic chelated Alizarin, respectively (in **Figure S15**).

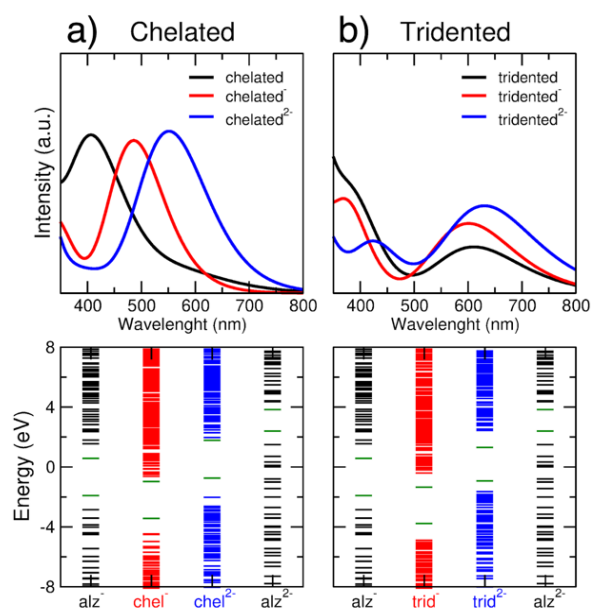

**Figure 8.** Top: Absorption spectra (B3LYP) of chelated (a) and tridented (b) monoanionic and dianionic complexes on $Cl_6$. Bottom: molecular orbital energies for the complexes reported together with those for the free monoanionic and dianionic Alizarin.

### 3.4.4 Alizarin/NP complex

In this section we investigate the optical properties of the Alizarin bonded to a realistic 2.2 nm NP. Due to the size and the number of atoms of this system, for the calculation of the spectra we use the PBE functional, which we have proved in the previous sections for the smaller $Cl_6$ and $Cl_{40}$ clusters, to describe with a satisfactory accuracy the main characteristics of the charge injection mechanism observed with the B3LYP functional. In **Figure 9a** we present the spectra of the Alizarin attached to the NP in the three different modes: tridented, chelated and bidented. The characteristics of each spectrum are similar to those observed for $Cl_{40}$: the tridented mode starts to absorb with an intense peak at 1.23 eV, while the chelated one has not an intense peak, but it presents an initial low intensity absorption feature. Finally, the bidented mode starts to absorb after the tridented and chelated mode as evidenced by an intense band at 1.67 eV. The analysis of the transitions provides valuable information for the understanding of the mechanism of charge injection. For the tridented mode the main excitation in the peak at 1.23 eV is due to the HOMO→LUMO+13 transition. The analysis of these orbitals (**Figure S16 d**) shows that both have similar characteristics to those of the HOMO and LUMO of free Alizarin, respectively, indicating that the peak at 1.23 eV corresponds to a red-shift in the absorption spectrum of the Alizarin, which depicts an indirect charge transfer mechanism. On the other hand, for the chelated configuration, the low-intense peak at 1.29 eV is mainly due to the HOMO→LUMO+8 transition. In this case the LUMO+8 orbital is centered on the NP. Although we carried out a careful search of the virtual orbitals involved in the excitations of the first part of the spectrum, we could not find an orbital with similar characteristics to those of the isolated Alizarin LUMO's orbital or centered on the molecule. As in the $Cl_6$ and $Cl_{40}$, this fact indicates that for the chelated adsorption mode a direct mechanism governs the charge injection. A similar direct mechanism is also found for the bidented adsorption mode. In the band centered at 1.67 eV the main excitation is due to the HOMO→LUMO+48 transition, where the HOMO orbital has the characteristics of the HOMO-1 orbital of the isolated molecule, while the LUMO+48 is localized mainly on the NP.

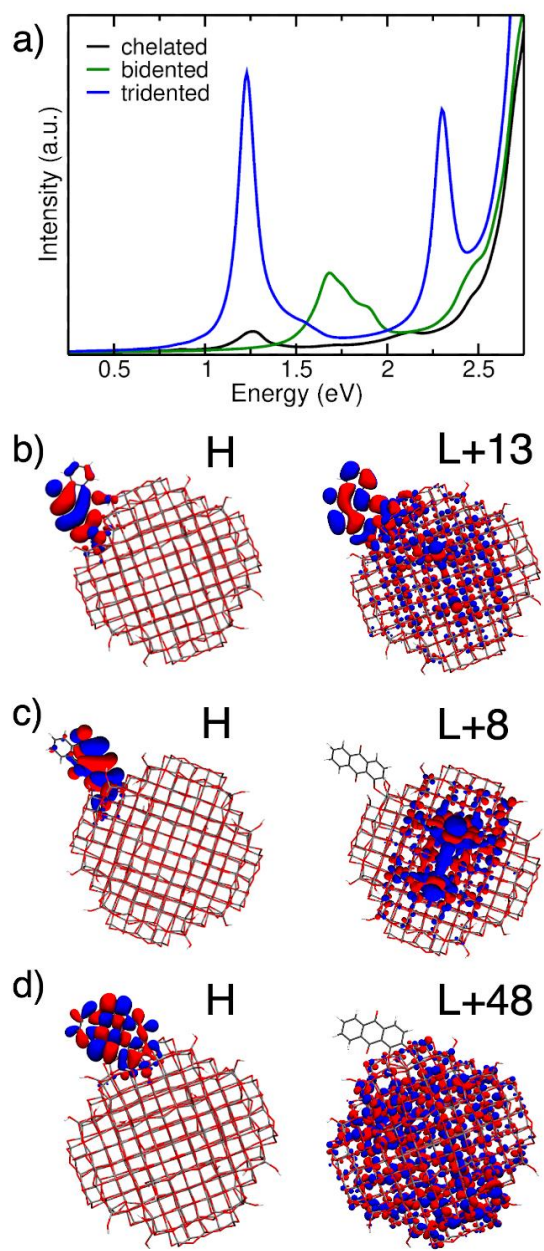

**Figure 9.** a) Absorption spectra (PBE) of tridented, chelated and bidented complexes on NP together with the Alizarin spectrum. b), c) and d) Selected occupied and virtual molecular orbitals of adsorbed Alizarin in tridented, chelated and bidented mode respectively.

In summary, we found that for the realistic NP the mechanisms are the same as those found for the $Cl_6$ and $Cl_{40}$ cluster models. It means that the type of charge injection mechanism (type I or type II) is determined by the type of adsorption of Alizarin on $TiO_2$. These results are very interesting, because by tuning the shape of the NP it is possible to favour or disfavour one mode of adsorption and, in this way, to control which injection mechanism will be at work. For instance, on spherical small NPs, where different types of uncoordinated Ti atoms ($Ti_{5c}$, $Ti_{4c}$, $Ti_{3c}OH$, $Ti_{4c}OH$), steps, and edges are present, the tridented adsorption mode will be feasible, introducing an indirect mechanism of charge injection. On the contrary, on faceted or big NPs, where most of the uncoordinated surface Ti atoms are $Ti_{5c}$ and only very few $Ti_{4c}$ atoms on the corners are present [66], the bidented mode is the most probable to take place, leading to a direct mechanism.

As mentioned in the Introduction, two electron injection mechanisms can be conceived. One is the indirect injection mechanism, which can be inferred from the similarity between the absorption spectrum of free molecule and that of the complex, and usually takes place in the non-adiabatic regime, i.e., electron tunneling through the potential barrier at the heterointerface between the dye and the semiconductor. The other is based on a direct photoexcitation of electron from the adsorbed molecule to the $TiO_2$ conduction band, which is suggested by the appearance of new low-energy absorption bands. Intermediate regimes between purely indirect and direct injection mechanisms can also coexist, for example in the case of *cis*-$(NCS)_2$-Ru(II)-bis(2,2'- bipyridine-4,4'-dicarboxylate)] dye adsorbed onto a $TiO_2$ model, different injection mechanisms were observed depending on the protonation state of the dye [35].

In a previous experimental work [29] multiple injection times were reported for the Alizarin/$TiO_2$ complexes, which is a hint for a non-adiabatic electron transfer, in contrast to what suggested by other works reporting an adiabatic electron transfer [22–24]. Experimental reports of the multiexponential injection kinetics have been explained in different ways: singlet and triplet state injection events [67], dye aggregation and multilayer formation on the surface [68], solvent effect on interfacial dynamics [69]. In Alizarin/$TiO_2$ complexes, however, the multiple injection time was previously rationalized with the discreteness of the conduction band levels caused by the finite size of the NPs used in the experiments (*r*~1.7nm) [29]. In this study, we have shown that an alternative way to explain the differences in the injection mechanism is to consider different modes of Alizarin binding to the surface. Since we can estimate that in the NPs used in these experiments [29] the percentage of $Ti_{4c}$ and $Ti_{5c}$ surface sites should be ~24% and ~32%, respectively (see Table V in Ref. [40]), therefore, we suggest that the experimental observation of different injection times [29] is due to different injection mechanisms for different adsorption configurations of Alizarin molecules on the various types of surface sites on the $TiO_2$ NPs.

## 4. Conclusions

In this work, we have presented a DFT (PBE and/or B3LYP) investigation of the different adsorption modes of Alizarin on $TiO_2$ clusters of different sizes (ranging from 6, to 40 to 223 $TiO_2$ units) and a TDDFT (PBE and/or B3LYP) study of the optical properties of the resulting Alizarin/$TiO_2$ complexes, with some test calculations also with the range-separated CAM-B3LYP functional.

First, we carried out a thorough investigation of the possible adsorption modes of Alizarin on $Cl_6$ and $Cl_{40}$ clusters and on the 2.2 nm NP. We found that Alizarin can bind to the $TiO_2$ surface in three different modes: tridented, bidented and chelated, in agreement with Raman spectroscopy experiments [30]. The tridented adsorption mode leads the most stable configurations, followed by the bidented and chelated ones.

Secondly, we performed a TDDFT investigation for the Alizarin molecule and the different $TiO_2$ models, in order to identify the best functional to describe the optical absorption spectrum of the separated components. B3LYP results are in very good agreement with experiments for both Alizarin and $TiO_2$. PBE functional underestimates the absorption onsets, but still correctly reproduces the characteristics of the spectra and of the transitions involved. Therefore, considering the computational cost, we used both B3LYP and PBE functionals for the Alizarin/$TiO_2$ complexes with $Cl_6$ and $Cl_{40}$ and only PBE functional for the Alizarin/$TiO_2$ NP.

Regarding the Alizarin/$TiO_2$ complexes, we found that, depending on the adsorption mode, the characteristics of absorption spectra change as well as the electron injection mechanism. We found that the experimental spectrum [25] for NPs of about 16 nm nicely correlates with the calculated spectrum for the bidented adsorption mode, which suggests that this type of binding is predominant on NPs of this size. From the analysis of the molecular orbitals, it was possible to determine that, for the tridented configuration, an indirect mechanism of injection is at play, whereas, for the bidented and chelated ones, a direct injection mechanism holds. It means that by tailoring the shape of the NP it is possible to get a desired injection mechanism.

Besides the adsorption mode of Alizarin to the $TiO_2$ surface, pH is also a crucial parameter. When we consider different proton states of the Alizarin/$TiO_2$ system (neutral, monoanionic and dianionic), which would result at increasing pH conditions, we observe a similar trend to that in the experiments (progressive red-shift) [26,64] for the chelated adsorption mode (not for the tridented one), especially when the water solvent is introduced, although in a simplified approach with an implicit model. Notably, going from the neutral to the anionic we also observe a change from a purely direct to a purely indirect electron injection mechanism.

Finally, the detailed results of this study help to unravel some theoretical and experimental controversies about the different types of electron injection mechanism (indirect [32,36,38] or direct [39]) and the multiple electron injection with different time scales reported on NPs of ~3.4 nm of diameter [29].

## Acknowledgements

The authors are grateful to Lorenzo Ferraro for his technical help. The project has received funding from the European Research Council (ERC) under the European Union's HORIZON2020 research and innovation programme (ERC Grant Agreement No [647020]).

**Data availability**

The raw/processed data required to reproduce these findings cannot be shared at this time as the data also forms part of an ongoing study.

Supplementary Material

# Tuning the electron injection mechanism by changing the adsorption mode: the case study of Alizarin on TiO$_2$

Federico Soria, Chiara Daldossi, Cristiana Di Valentin[*]

Dipartimento di Scienza dei Materiali, Università di Milano Bicocca
Via R. Cozzi 55, 20125 Milano Italy

[*] Corresponding author: cristiana.divalentin@unimib.it

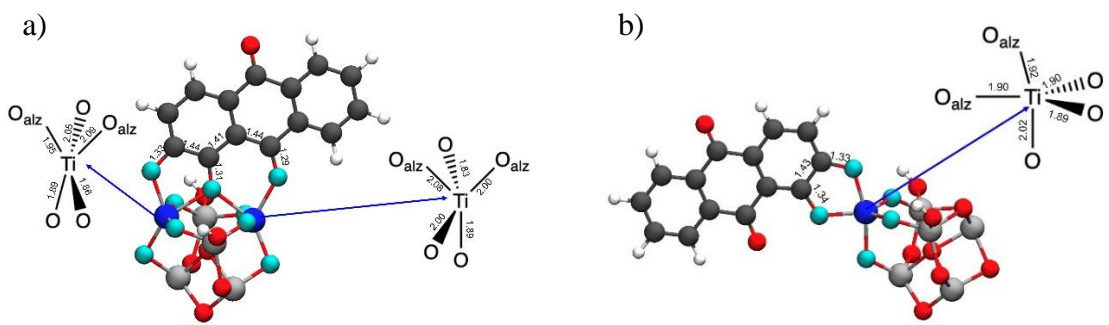

**Figure S1.** Optimized structures, with the distance details in Å, of the adsorption of Alizarin in a) tridented, and b) chelated modes on $Cl_6$ cluster.

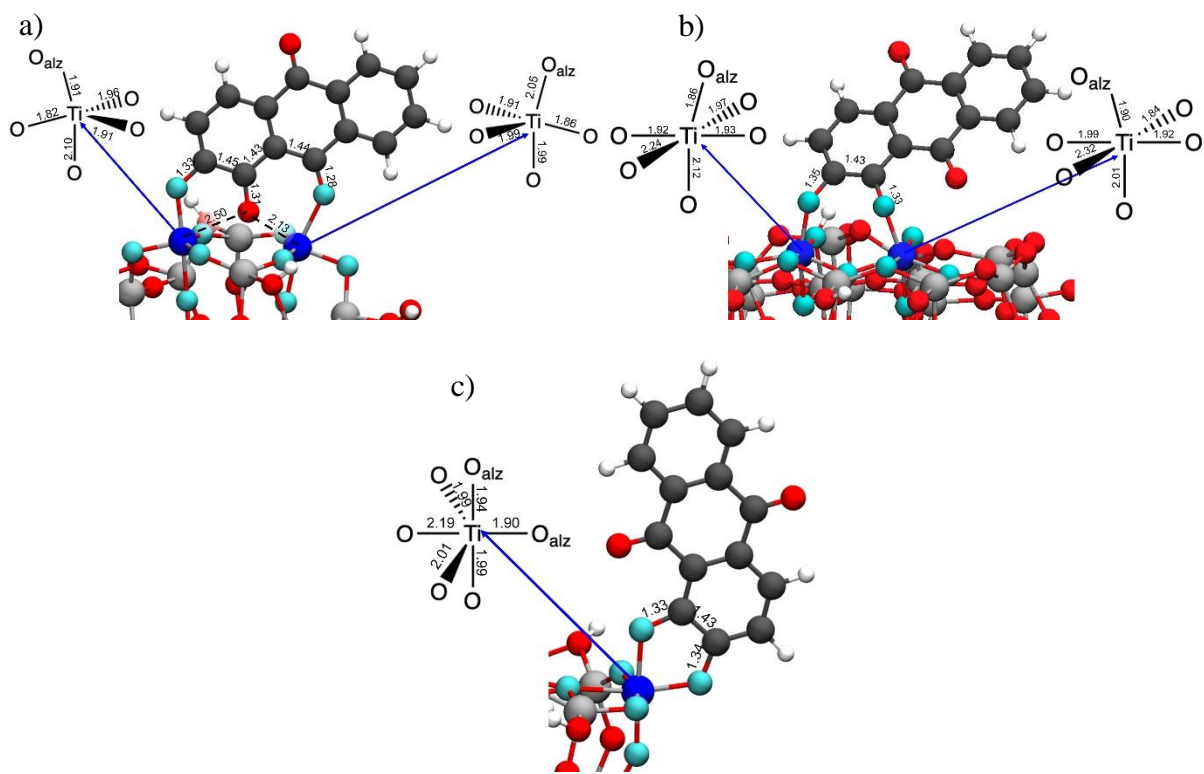

**Figure S2.** Optimized structures, with the distance details in Å, of the adsorption of Alizarin in a) tridented, b) bidented and c) chelated modes on $Cl_{40}$ cluster.

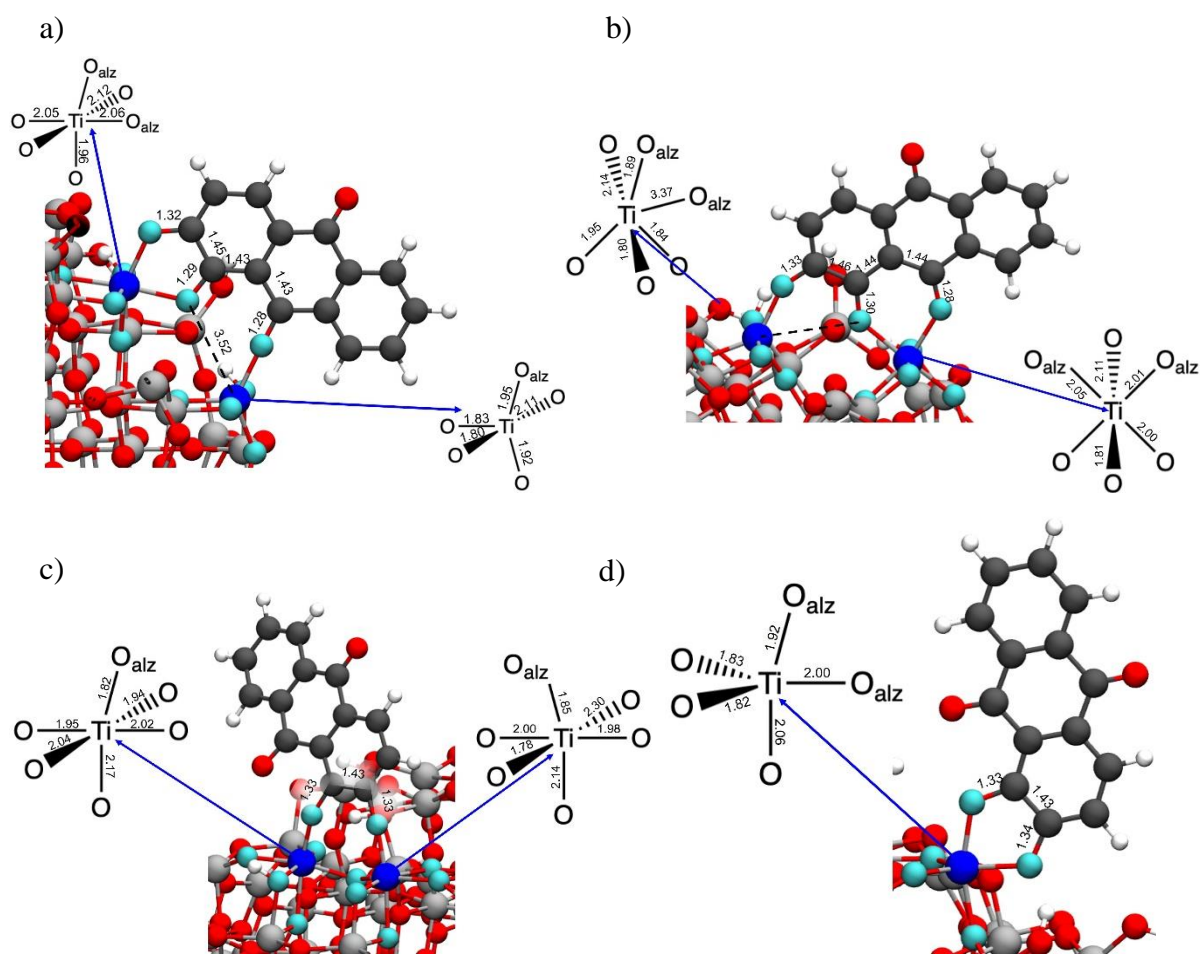

**Figure S3.** Optimized structures, with the distance details in Å, of the adsorption of Alizarin in a) and b) tridented, c) bidented and d) chelated modes on 2.2 nm NP.

**Table S1.** Calculated binding energies in eV for the adsorption of Alizarin in the different configurations on the Cl$_6$ and Cl$_{40}$ TiO$_2$ models and a realistic NP using PBE functional.

| a) Cl$_{40}$/PBE | E$_{ads}$ | E$_{def-clus}$ | E$_{def-aliz}$ | E$_{bind}$ |
|---|---|---|---|---|
| Trid | **-0.18** | 6.12 | 0.21 | -8.46 |
| Bid | **-0.43** | 4.07 | 0.14 | -7.32 |
| chel | **0.04** | 5.08 | 0.25 | -7.05 |

| b) NP/PBE | E$_{ads}$ | E$_{def-clus}$ | E$_{def-aliz}$ | E$_{bind}$ |
|---|---|---|---|---|
| Trid | **-3.48** | 3.91 | 0.41 | -8.73 |
| Bid | **-1.46** | 3.80 | 0.28 | -7.30 |
| Chel | **-2.12** | 3.62 | 0.38 | -8.33 |

| c) Cl$_6$/PBE | E$_{ads}$ | E$_{def-clus}$ | E$_{def-aliz}$ | E$_{bind}$ |
|---|---|---|---|---|
| Trid | **-2.99** | 5.23 | 0.40 | -9.46 |
| Chel | **-2.38** | 3.62 | 0.30 | -8.05 |

E$_{ads}$: the adsorption energy of Alizarin on TiO$_2$ (E$_{complex}$-E$_{TiO2}$-E$_{Alizarin}$)

E$_{def-clus}$: the distortion energy of the TiO$_2$ in the complex (E$_{TiO2\ in\ the\ geometry\ of\ complex}$-E$_{TiO2}$)

E$_{def-aliz}$: the distortion energy of Alizarin (E$_{Alizarin\ in\ the\ geometry\ of\ complex}$-E$_{Alizarin}$)

E$_{bind}$: the binding energy, evaluated as the energy difference between the complex and the separated fragments (Alizarin and cluster/NP) in the same geometry as in the complex (E$_{complex}$-E$_{TiO2(+2H)}$-E$_{alizarin(-2H)}$) after homolytic dissociation (spin polarization is taken into account).

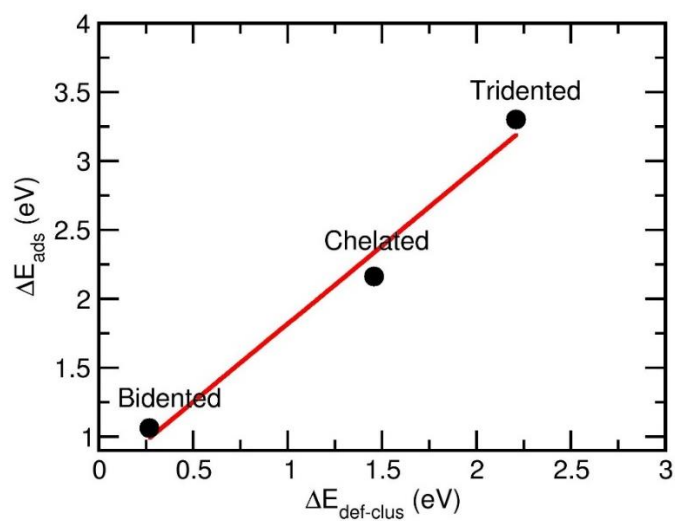

**Figure S4.** Difference in the adsorption energies between Cl$_{40}$ and NP ($\Delta$E$_{ads}$) as a function of the difference in the deformation energies ($\Delta$E$_{def-clus}$) for each configuration.

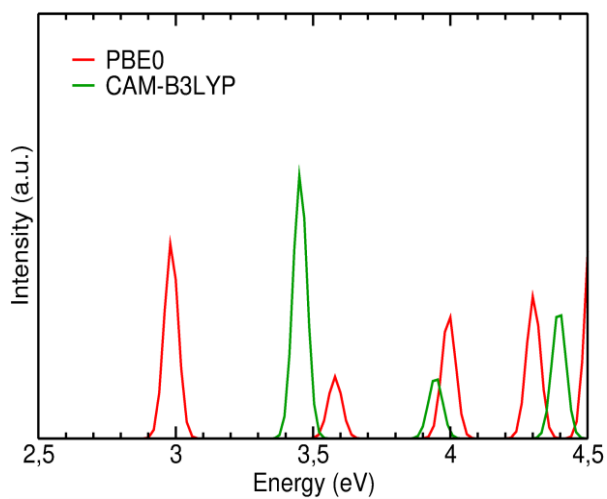

**Figure S5.** Optical spectra of the Alizarin molecule calculated in vacuum using PBE0 and CAM-B3LYP functionals.

**Table S2:** Lowest band energies (eV) and HOMO-LUMO gap (eV) for free Alizarin using the PBE, PBE0, B3LYP and CAM-B3LYP functionals, obtained from LR-TDDFT and DFT calculations respectively.

|                 | PBE  | PBE0 | B3LYP | CAM-B3LYP | exp  |
|-----------------|------|------|-------|-----------|------|
| First excitation | 2.31 | 2.98 | 2.86  | 3.45      | 2.88 |
| HOMO-LUMO gap   | 1.88 | 3.64 | 3.30  | 5.83      |      |

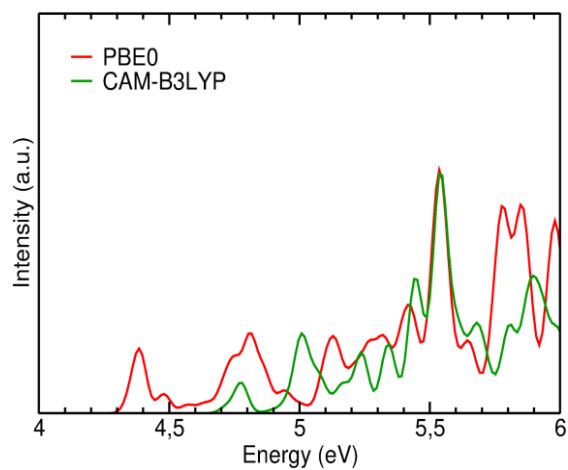

**Figure S6.** Optical spectrum of the Cl$_6$, calculated by the use of PBE0 and CAM-B3LYP functionals.

**Table S3.** Lowest band energies (eV) and HOMO-LUMO gap (eV) for Cl$_6$ using the PBE, PBE0, B3LYP and CAM-B3LYP functionals, obtained from LR-TDDFT and DFT calculations respectively.

|  | Cl$_6$ | | | |
|---|---|---|---|---|
|  | PBE | PBE0 | B3LYP | CAM-B3LYP |
| First excitation | 2.68 | 4.39 | 4.01 | 4.75/5.00 |
| HOMO-LUMO gap | 2.66 | 5.31 | 4.70 | 7.91 |

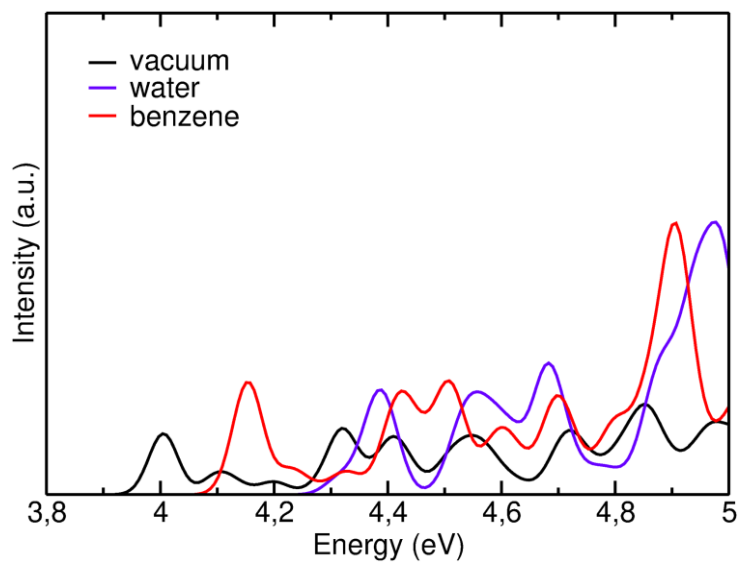

**Figure S7.** Optical spectrum of $Cl_6$ calculated in vacuum, and in water and benzene solvents using B3LYP functional.

**Table S4.** Lowest band energies (eV) and HOMO-LUMO gap (eV) for $Cl_6$ using the B3LYP functional, in vacuum, water and benzene, obtained from LR-TDDFT and DFT calculations respectively.

| B3LYP | vacuum | water | benzene |
| --- | --- | --- | --- |
| First excitation | 4.01 | 4.38 | 4.16 |
| HOMO-LUMO gap | 4.70 | 5.16 | 4.87 |

**Table S5**. Assignments of B3LYP electronic excitations for Alizarin/TiO$_2$ complexes formed on the Cl$_6$ cluster. For each transition, the main contribution is marked in bold.

| transition energy (eV), excitation number | oscillator strength | wave function |
|---|---|---|
| **Tridented** | | |
| 2.03 (1) | 0.0771 | **H →L (97.4%)** |
| 2.64 (3) | 0.0058 | **H→L+2 (74.4%)**, H→L+4 (19.1%) |
| 2.96 (7) | 0.0371 | **H-1→L (57.3%)**, H→L+1 (4.7%), H→L+2 (3.3%), H→L+3 (6.1%), H→L+4 (8.2%), H→L+5 (12.9%), H→L+7 (3.0%) |
| 3.10 (8) | 0.0411 | H-1→L (11.9%), H→L+3 (6.1%), **H→L+5 (58.0%)**, H→L+6 (4.8%), H→L+7 (5.7%), H→ L+9 (3.1%), H →L+10 (3.0%) |
| **chelated** | | |
| 2.08 (4) | 0.0180 | H→L+1 (2.90%), **H→L+3 (32.7%), H→L+4 (37.2%)**, H→L+5 (20.8%) |
| 2.21 (5) | 0.0101 | **H→L+2 (83.5%)**, H→L+3 (8.9%), H→ L+4 (5.8) |
| 2.52 (8) | 0.0155 | H-1→L (16.2%), H→L+4 (2.1%), **H→L+6 (77.5%)** |
| 2.75 (13) | 0.0773 | H→L+8 (7.0%), **H→L+9 (75.0%),** H→L+5 (2.3%) |

**Table S6**. Assignments of B3LYP electronic excitations for Alizarin/TiO$_2$ complexes formed on the Cl$_{40}$ cluster. For each transition, the main contribution is marked in bold.

| transition energy (eV), excitation number | oscillator strength | wave function |
|---|---|---|
| **Tridented** | | |
| 1.87 (1) | 0.0490 | **H→L (86.4%),** H→L+1 (4.8%), H→L+2 (4.2%) |
| 2.21 (4) | 0.0105 | H→L (7.8%), **H→L+1 (22.2%), H→L+2 (57.0%)**, H→L+7 (6.29%) |
| 2.37 (5) | 0.0171 | **H→L+3 (23.8%),** H→L+6 (12.7%), H→L+7 (9.0%), H→L+10 (15.3%), H→L+11 (5.8%), H→L+12 (10.0%), |
| 2.45 (7) | 0.0191 | H→L+1(5.7%), **H→L+4 (22.1%),** H→L+5 (16.6%), **H→L+13 (21.5%),** |
| 2.68 (14) | 0.0553 | H→L(7.0%), **H→L+8 (25.8%)**, H→L+9 (13.1%), H→L+23 (6.3%), |
| **chelated** | | |
| 1.36 (1) | 0.0015 | **H→L (90.6%)**, H→L+1 (6.0%) |
| 1.46 (2) | 0.0031 | H→L (6.5%), **H→L+1 (89.4%)** |
| 2.02 (15) | 0.0071 | **H→L+10 (18.0%)**, H→L+11 (6.7%), H→L+13 (6.4%), H→L+16 (8.0%), H→L+23 (7.0%), |
| 2.34 (40) | 0.0325 | H→ L+27 (5.7%), **H→L+29 (12.4%)**, H→L +33 (6.6%), **H→L+34 (13.7%)**, H→L+37 (9.3%), H→L+46 (6.9%) |
| 2.61 (67) | 0.0196 | H→L+47 (9.0%), **H→L+48 (11.8%)**, **H→L+53 (12.4%)**, H→L+54 (7.4%), H→L+55 (6.7%), **H→L+62 (10.4%),** |
| **bidented** | | |
| 2.38 (2) | 0.0469 | H →L (14.9%), **H→ L+1 (36.6%)**, H→ L+2 (2.9%), **H→ L+3 (22.8%),** H→ L+5 (6.8%) |
| 2.42 (3) | 0.0489 | H →L (9.2%), H→ L+1 (12.4%), **H→ L+2 (29.1%)**, **H→ L+3 (34.9%)** |
| 2.82 (14) | 0.0153 | **H-1→L+1 (14.8%)**, H→L+7 (9.7%), H→L+8 (5.5%), H→L+16 (6.9%), **H→L+17 (20.8%)**, H→L+21 (9.6%) |
| 3.00 (26) | 0.0314 | H→L+16 (7.8%), **H→L+19 (13.5%)**, H→L+24 (5.7%), **H→L+26 (10.2%)**, H→L+28 (10.4%), H→L+32 (6.1%) |

**Table S7.** Indices of Charge Transfer Distance, $D_{CT}$ and Charge Passed, $q_{CT}$, for the most important transition for Alizarin isolated attached on $Cl_{40}$ and $Cl_6$ in the different adsorption modes

| System | w(eV) | $D_{CT}$ (Å) | $q_{CT}$ (a.u.) |
|---|---|---|---|
| Alizarin | 2.864 | 3.556 | 0.753 |
| $Cl_{40}$+trid | 1.869 | 3.678 | 0.810 |
| $Cl_{40}$+bid | 2.375 | 4.510 | 0.852 |
| $Cl_{40}$+chel | 1.368 | 9.318 | 0.978 |
| $Cl_6$+chel | 1.798 | 7.009 | 0.976 |
| $Cl_6$+trid | 2.027 | 3.247 | 0.700 |

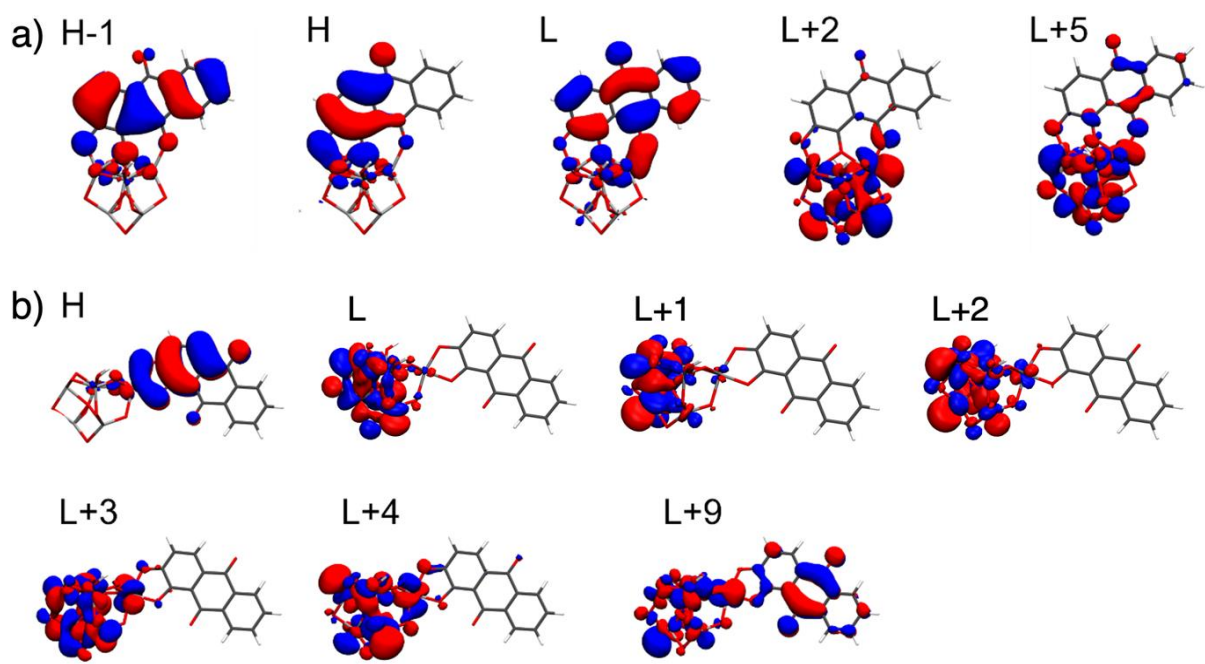

**Figure S8.** Selected virtual molecular orbitals of adsorbed Alizarin in a) tridented and b) chelated mode.

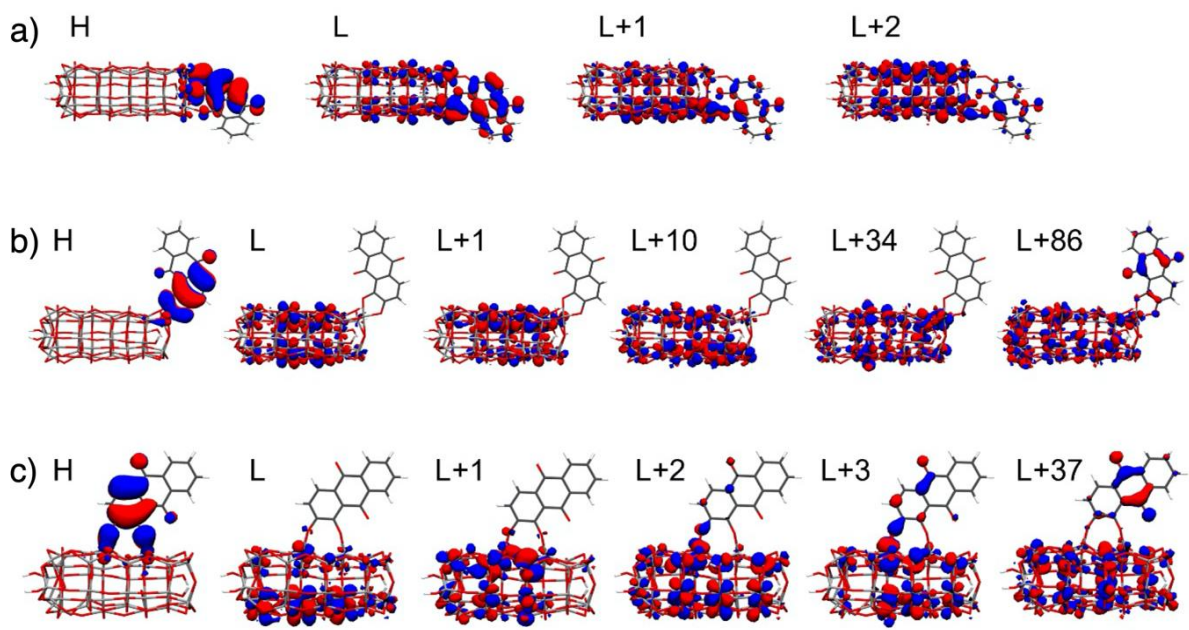

**Figure S9.** Selected virtual molecular orbitals of adsorbed Alizarin in a) tridented, b) chelated and c) bidented mode.

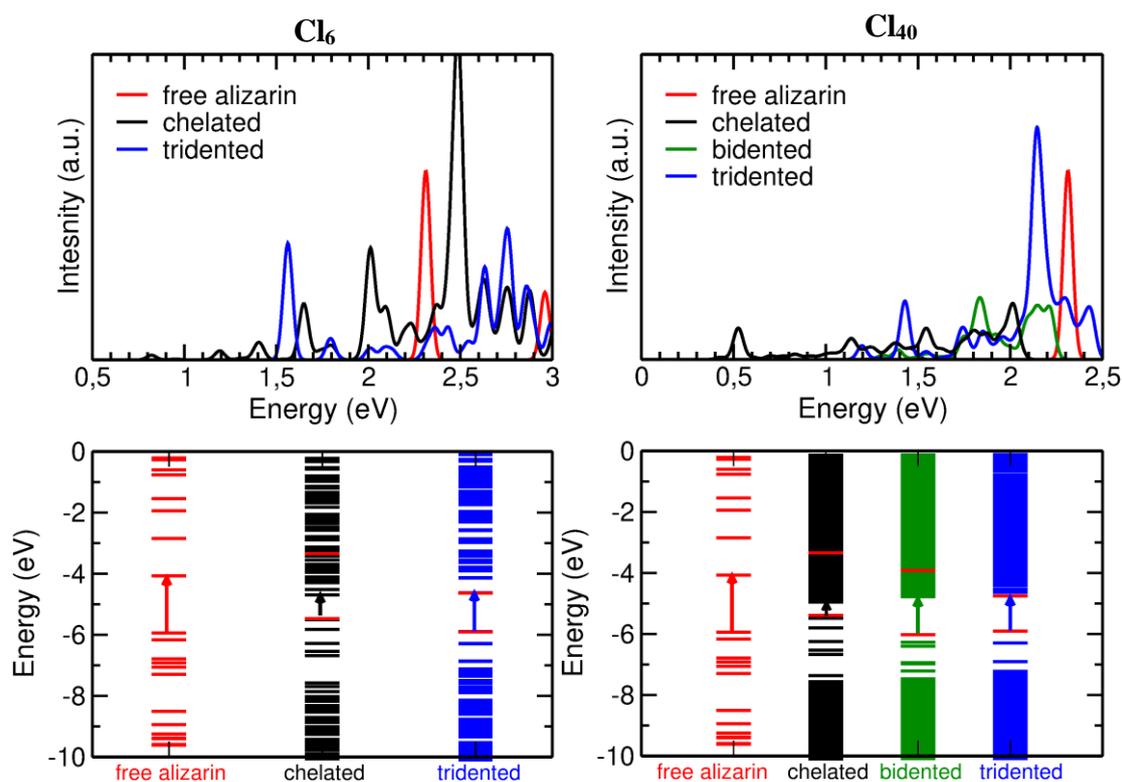

**Figure S10.** Top: absorption spectra of complexes formed on $Cl_6$ and $Cl_{40}$ respectively together with the Alizarin spectrum using the PBE functional. Bottom: Molecular orbital energies for the ground state of free Alizarin, and the complexes on $Cl_6$ and $Cl_{40}$ respectively. For each complex we have marked the HOMO orbital and the orbital that corresponds to the LUMO of the free molecule.

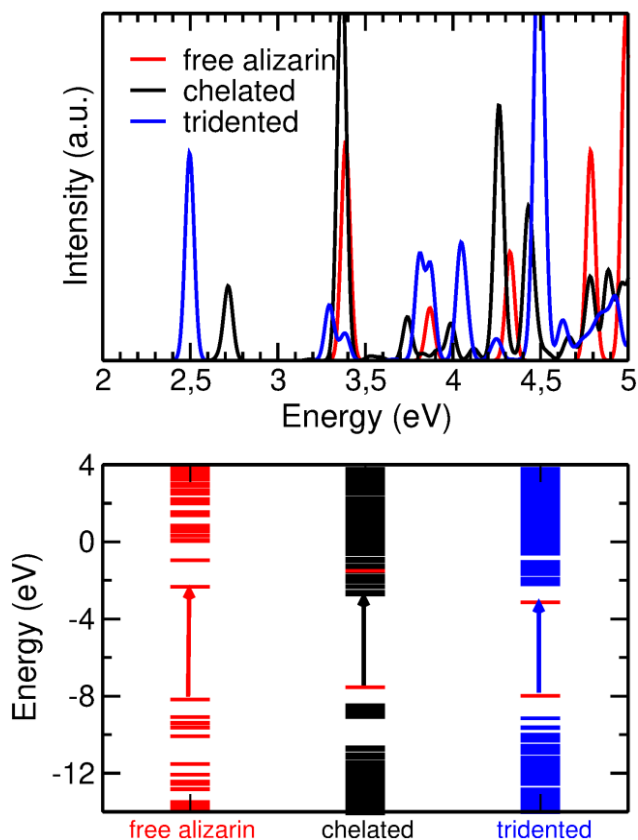

**Figure S11.** Top: absorption spectra of complexes formed on $Cl_6$ together with the Alizarin spectrum using the CAM-B3LYP functional. Bottom: Molecular orbital energies for the ground state of free Alizarin, and the complexes on $Cl_6$. For each complex we have marked the HOMO orbital and the orbital that corresponds to the LUMO of the free molecule.

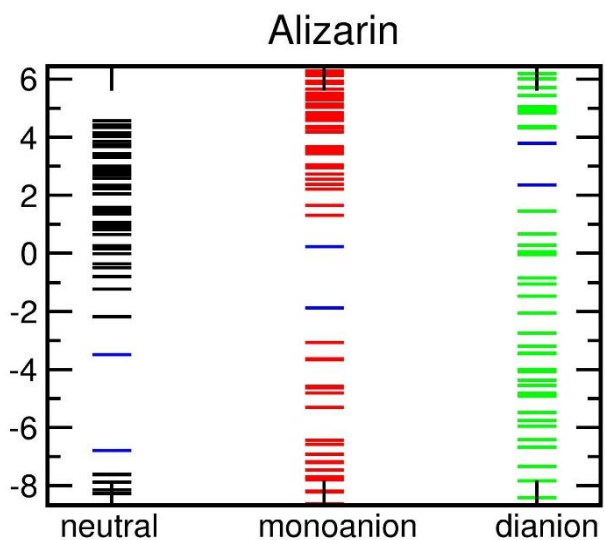

**Figure S12.** Molecular orbital energies for the ground state of free Alizarin, neutral, monoanionic and dianionic respectively using B3LYP hybrid functional. We have marked the HOMO and LUMO orbitals for each specie.

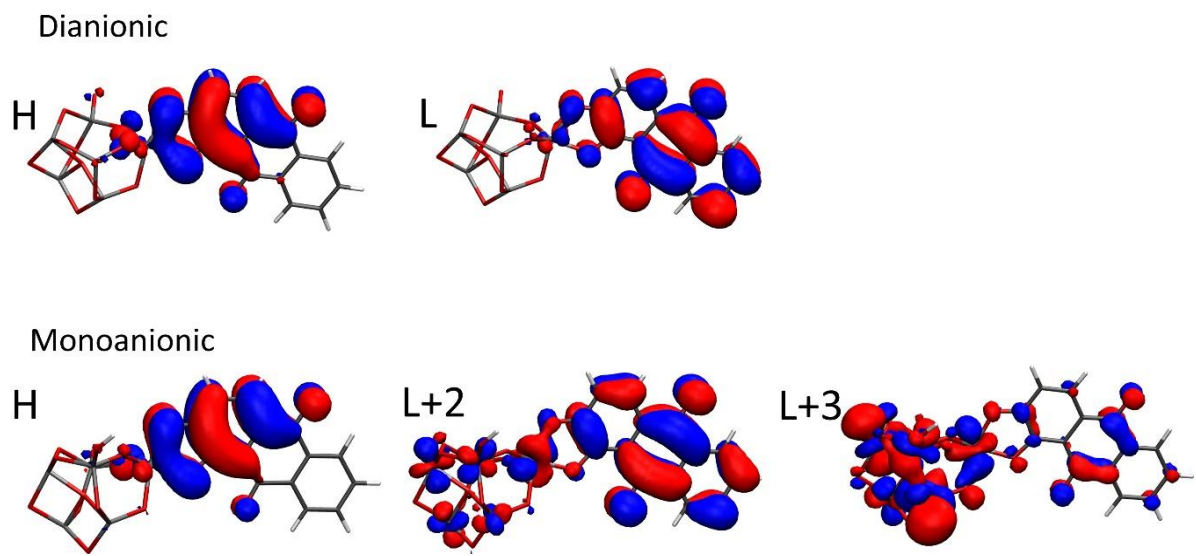

**Figure S13.** Selected virtual molecular orbitals of charged adsorbed Alizarin in chelated mode. Top: Dianionic System. Bottom: Monoanionic System.

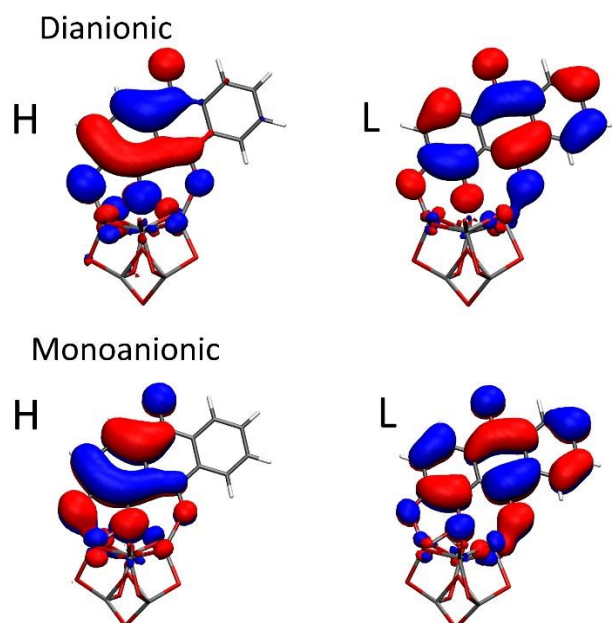

**Figure S14.** Selected virtual molecular orbitals of charged adsorbed Alizarin in tridented mode. Top: Dianionic System. Bottom: Monoanionic System.

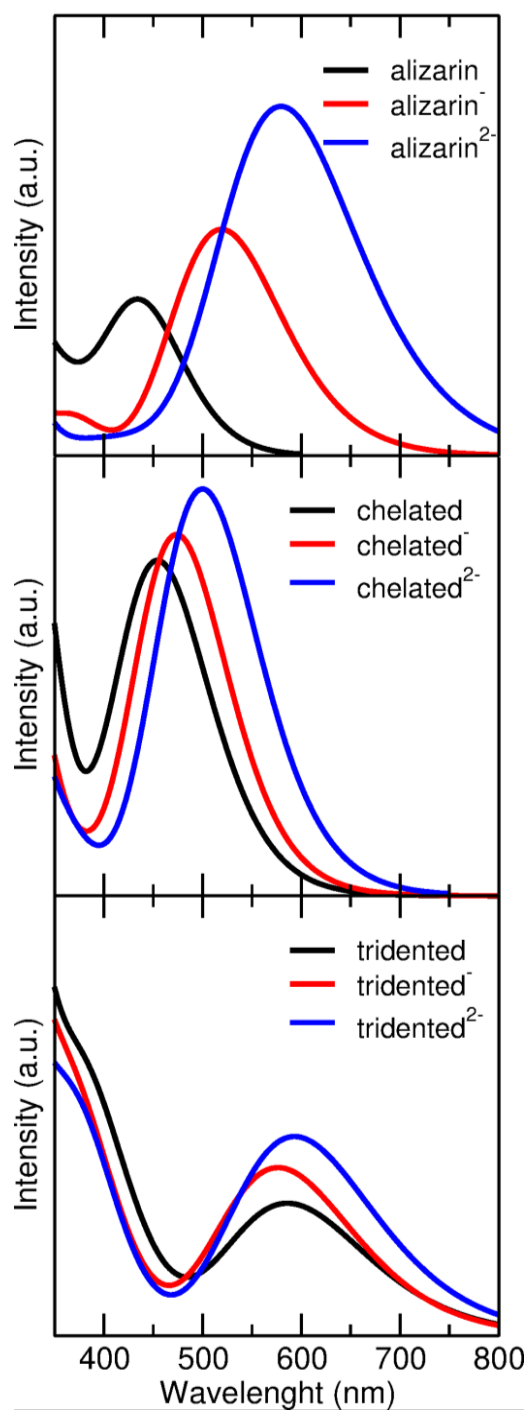

**Figure S15.** Simulated optical spectra of neutral and charged systems in implicit water as a solvent. The inclusion of solvent was done by use of the PCM model. Top: free Alizarin. Middle: Alizarin attached on $TiO_2$ $Cl_6$ in a chelated mode. Bottom: Alizarin attached on $TiO_2$ $Cl_6$ in a tridented mode.

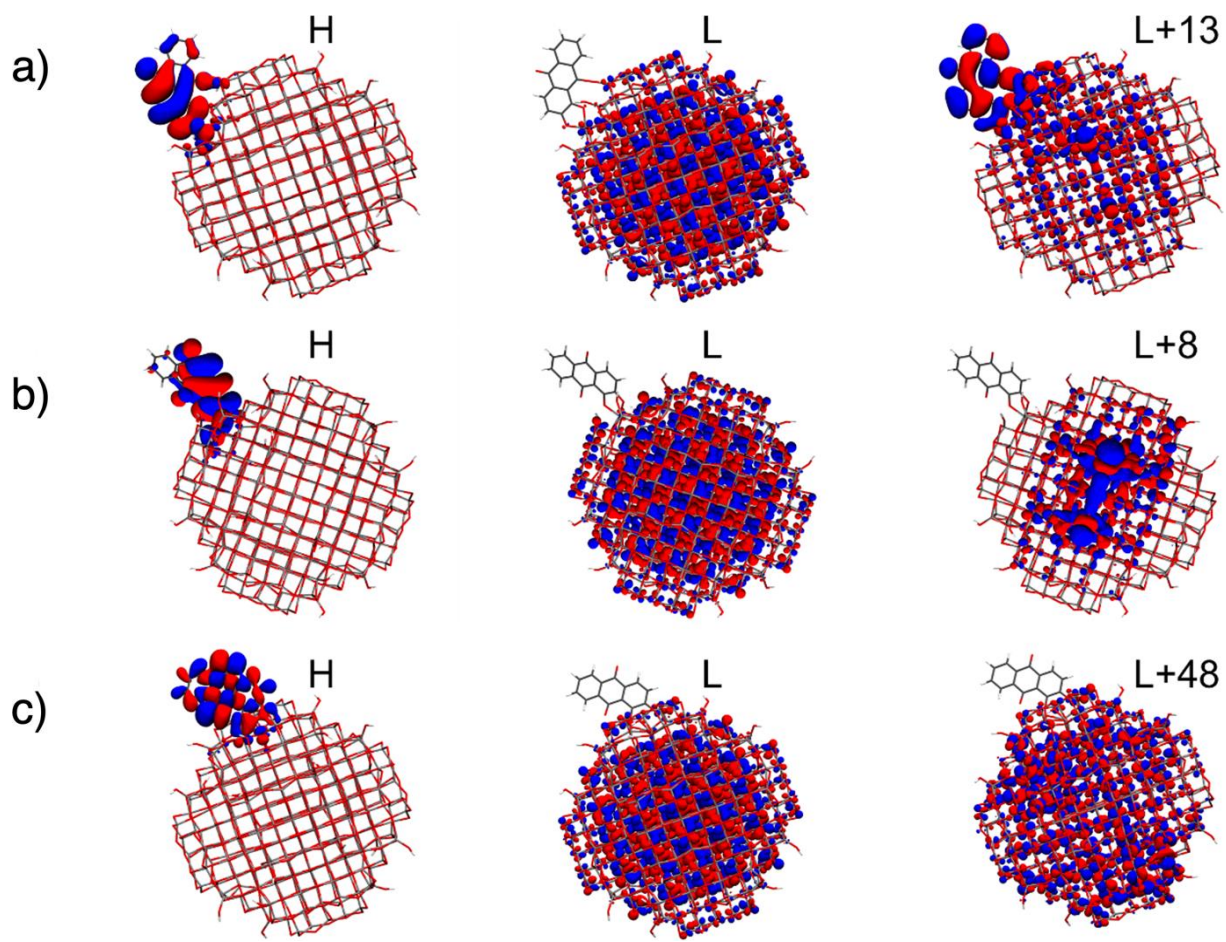

**Figure S16.** Selected virtual molecular orbitals of adsorbed Alizarin in a) tridented, b) chelated and c) bidented mode.